\documentclass[a4paper,11pt]{article}
\pdfoutput=1 

\usepackage{jcappub}
\usepackage[toc,page]{appendix}
\usepackage{subcaption}
\usepackage{enumerate}
\usepackage{amsmath}

\usepackage{graphicx}                
\usepackage{multirow}
\usepackage{braket}
\usepackage[T1]{fontenc} 
\usepackage[table]{xcolor}
\usepackage{float}
\def\noi{{\noindent}}

\definecolor{verde}{rgb}{0,0.5,0}

\def\be{\begin{equation}}
\def\ee{\end{equation}}
\def\bea{\begin{eqnarray}}
\def\eea{\end{eqnarray}}
\def\be{\begin{equation}}
\def\ee{\end{equation}}
\def\ba{\begin{eqnarray}}
\def\ea{\end{eqnarray}}

\hypersetup{pdftitle={ Revisiting 1-loop corrections in single-field inflation}
}
\usepackage{bm}

\title{\boldmath Numerical 1-loop correction from a potential yielding ultra-slow-roll dynamics}

\author[a]{Matthew W. Davies,}
\author[a,b]{Laura Iacconi}
\author[a]{and David J. Mulryne}
\affiliation{$^{a}$Astronomy Unit, Queen Mary University of London, \\
Mile End Road, London, E1 4NS, UK}
\affiliation{$^{b}$Institute of Cosmology \& Gravitation, University of Portsmouth, \\
Burnaby Road, Portsmouth, PO1 3FX, UK}

\emailAdd{m.w.davies@qmul.ac.uk}
\emailAdd{l.iacconi@qmul.ac.uk}
\emailAdd{d.mulryne@qmul.ac.uk}

\abstract{Single-field models of inflation might lead to amplified scalar fluctuations on small scales due, for example, to a transient ultra-slow-roll phase. It was argued by Kristiano \& Yokoyama in Ref.~\cite{Kristiano:2022maq} that the enhanced amplitude of the scalar power spectrum on small scales has the potential to induce a sizeable 1-loop correction to the spectrum at large scales. In this work, we repeat the calculation for the 1-loop correction presented in Ref.~\cite{Kristiano:2022maq}. We closely follow their assumptions but evaluate the loop numerically. 
This allows us to consider both instantaneous and smooth transitions between the slow-roll and ultra-slow-roll phases.
In particular, we generate models featuring realistic, smooth evolution from an analytic inflationary potential.
We find that, upon fixing the amplitude of the peak in the power spectrum at short scales, 
the resulting 1-loop correction is not significantly reduced by considering a smooth evolution. 
In particular, for a power spectrum with a tree-level peak amplitude 
potentially relevant for small-scale phenomenology, 
e.g. primordial black hole production, 
the 1-loop correction on large scales is a few percent of the tree-level power spectrum.
}

\begin{document}
	\maketitle
	\flushbottom
	
\section{Introduction}
\label{sec: intro}
\noi In recent years, there has been considerable interest in models of 
inflation that are capable of setting the initial conditions needed to produce a large population of primordial black holes (PBHs)~\cite{Carr:1974nx,Carr:2020gox}. 
This interest has been driven by many factors, including the possibility that PBHs could contribute to at least a fraction of the black-hole merger events detected by the LIGO-Virgo-KAGRA collaboration (see e.g. Ref.~\cite{Franciolini:2022tfm} for a recent investigation),
as well as the potential for PBHs to explain a fraction, or even the totality, of dark matter~\cite{Carr:2016drx,Bertone:2018krk, Bartolo:2018evs,Green:2020jor}.

Within the context of canonical, single-field inflation, producing PBHs usually requires
a self-interaction potential that supports three distinct phases of evolution. 
First, there must be a flat region of the potential yielding a period of slow-roll (SR) inflation. 
During this era nearly scale-invariant and approximately Gaussian scalar perturbations are seeded, in agreement with CMB observations on large scales ($0.005\,\text{Mpc}^{-1}\lesssim k\lesssim 0.2 \,\text{Mpc}^{-1}$)~\cite{Planck:2018jri, BICEPKeck:2021gln}. 
Next, there must be a subsequent region of the potential where the inflaton velocity decreases rapidly, e.g. a phase of ultra-slow roll (USR)~\cite{Kinney:2005vj, Dimopoulos:2017ged, Pattison:2018bct} or generic constant roll~\cite{Motohashi:2014ppa, Motohashi:2019rhu, Motohashi:2023syh}. 
This is possible, for example, when the potential features an almost-stationary inflection point~\cite{Garcia-Bellido:2017mdw, Germani:2017bcs, Ballesteros:2017fsr} or a superimposed bump (or dip)~\cite{Atal:2019cdz, Mishra:2019pzq}. 
Remarkably, the USR dynamics amplifies scalar fluctuations, yielding a peak in the scalar power spectrum on scales smaller than those probed by CMB experiments. 
Upon horizon re-entry during the radiation-dominated phase, these large perturbations may collapse to form PBHs~\cite{10.1093/mnras/168.2.399} (see e.g. the review~\cite{Sasaki:2018dmp}), as well as leading to large gravitational waves produced at second-order in perturbation theory~\cite{Ananda:2006af, Baumann:2007zm}.
In order to trigger PBH formation, the small-scale power spectrum needs to be enhanced roughly by seven orders of magnitude with respect to the large-scale value~\cite{Motohashi:2017kbs}. 
Finally, at the end of the USR phase, the system typically transitions back to SR dynamics, and inflation eventually ends. For brevity, we will label the three-phase model as SR-USR-SR.

Given the huge amplification necessary for PBH production,
it is important to assess whether a model can self-consistently produce both the seeds of large-scale structure and of PBHs.
In particular, an issue which has recently received considerable attention is that of loop corrections to the power spectrum on large scales\footnote{For studies on the contribution of enhanced small-scale models to the 1-loop power spectrum evaluated at the peak scales and at near infrared scales see Ref.~\cite{Iacconi:2023slv} and Ref.~\cite{ Fumagalli:2023loc} respectively.} 
due to the enhanced short-scale modes. 
Indeed, one might worry that due to amplified fluctuations at peak scales, the magnitude of the entire series of loop corrections could be larger than the tree-level contribution, i.e. the one computed using the linear equations of motion for the perturbations.
Large loop corrections could signal a breakdown of perturbativity for the underlying model. 
They would indicate that the tree-level power spectrum does not provide an accurate description for the power spectrum, and should therefore not be employed when comparing inflationary predictions with observations.
This is not desirable on large scales, where CMB measurements provide strict constraints on the power spectrum, in accordance with the predictions from single-field SR inflation. 

As a first step in investigating these issues, several authors focused on gauging the magnitude of the 1-loop correction (1LC), see Refs.~\cite{Inomata:2022yte,Cheng_2022, Kristiano:2022maq,Riotto:2023hoz, Choudhury:2023vuj,
Choudhury:2023jlt, Kristiano:2023scm, Riotto:2023gpm,
Firouzjahi:2023aum, Motohashi:2023syh, Choudhury:2023rks,
Choudhury:2023hvf, Firouzjahi:2023ahg, Firouzjahi:2023btw, Franciolini:2023lgy, Tasinato:2023ukp, Cheng:2023ikq,
Fumagalli:2023hpa, Maity:2023qzw, Tada:2023rgp,
Firouzjahi:2023bkt}. 
In particular, in Ref.~\cite{Kristiano:2022maq} Kristiano \& Yokoyama estimate a contribution to the 1LC produced within models featuring a USR phase.  
Their calculation is performed analytically under the assumption of instant transitions between the USR and SR phases. They find that, for models that could lead to PBH production, the 1LC is comparable to the tree-level contribution and therefore argue that PBH formation within single-field inflation is ruled out. 
One might, however, question the assumptions under which this calculation is performed, and whether they are the origin of the large 1LC found. 
First of all, instantaneous transitions are unphysical, and can lead to spurious effects such as oscillations in the power spectrum at peak scales, 
see e.g. Ref.~\cite{Cole:2022xqc}, and large non-Gaussianity on small-scales.
For example, in Ref.~\cite{Cai:2018dkf} Cai \textit{et al.} show that while a phase of USR yields $f_\text{NL}=5/2$~\cite{Namjoo:2012aa}, the subsequent smooth evolution from USR to SR characteristic of realistic models washes it away, yielding a SR-suppressed final value. 
The 1LC evaluated in Ref.~\cite{Kristiano:2022maq} is dependent on using the same term in the interaction Hamiltonian that leads to the production of large non-Gaussianity during USR. 
One assumption that should be put under scrutiny, therefore, is that of instantaneous transitions.  
In Ref.~\cite{Franciolini:2023lgy} Franciolini \textit{et al.} allow for more realistic, extended transitions and numerically evaluate the resulting 1LC.
Their results show that the 1LC on large scales is still appreciable, though not large enough to rule out these models for PBH formation, even for instantaneous transitions.
Moreover, the 1LC does not change significantly between scenarios with instant and non-instant USR to SR transitions. 
In their work, however, the background and perturbations evolution does not follow from an explicit inflationary potential, but rather the non-instant transitions are modelled using an analytic \textit{ansatz} for the time-evolution of the second SR parameter, $\eta$, see also Refs.~\cite{Byrnes:2018txb, Taoso:2021uvl, Franciolini:2022pav, Cole:2022xqc}.

In our work, we take a step forward and numerically evaluate the 1LC to the large-scale scalar power spectrum \textit{directly} from an explicit inflaton potential, $V(\phi)$, that supports USR dynamics. 
While allowing us to explore the case of smooth USR-SR transitions directly from an explicit model, 
our analysis also provides a framework which could be easily extended for the analysis of any single-field potential one might want to investigate.

\medskip
\textit{Content:} This work is structured as follows. In Sec.~\ref{sec:loop with instantaneous transitions} we review the main aspects of the calculation of Ref.~\cite{Kristiano:2022maq} and provide numerical results for the 1LC evaluated in models with instant USR-SR transitions.  
In Sec.~\ref{sec:loop from potential model}, we then proceed to discuss the 1LC produced from an explicit potential model. 
In particular, we introduce the model in Sec.~\ref{sec:potentialmod} and in Sec.~\ref{sec: the in in calculation} present an analytical expression for the 1LC calculated within the In-In formalism which is best suited for numerical evaluation. 
In Sec.~\ref{sec:compare} we compare the 1LC obtained for instantaneous and smooth transitions, which constitute the main results of this work. 
Additionally, we carefully test the robustness of our numerical analysis in Sec.~\ref{sec:robust}, investigating which regime of evolution contributes the most to the 1LC. We conclude in Sec.~\ref{sec: discussion}.

\medskip
\textit{Conventions:} Throughout this work, we consider a spatially-flat Friedmann--Lema\^{i}tre--Robertson--Walker universe, with line element $\text{d}s^2=-\text{d}t^2+a^2(t)\delta_{ij}\text{d}x^i\text{d}x^j$, where $t$ denotes cosmic time and $a(t)$ is the scale factor. The Hubble rate is defined as $H\equiv {\dot a}/{a}$,  where a derivative with respect to cosmic time is denoted by $\dot f \equiv {\mathrm{d}f}/{\mathrm{d}t}$. Conformal time is defined as $\mathrm{d}\tau \equiv \mathrm{d}t/a(t)$. The number of e-folds of expansion is defined as $N\equiv \int\,H(t)\mathrm{d}t$ and $f'\equiv {\mathrm{d}f}/{\mathrm{d}N}$. We use natural units and set the reduced Planck mass, $M_\text{Pl}\equiv(8\pi G_N)^{-1/2}$, to unity unless otherwise stated.

\section{1-loop correction on large scales: SR-USR-SR model with instantaneous transitions}
\label{sec:loop with instantaneous transitions}

Here we review the calculation of the 1-loop correction induced on large scales by enhanced, small-scale modes. We follow closely the work of Kristiano \& Yokoyama~\cite{Kristiano:2022maq} in which the transitions from the initial SR phase to USR and from USR to the final SR phase are modelled as being instantaneous.
This will allow us to clarify the differences between the analytic calculation of Ref.~\cite{Kristiano:2022maq} and our results in Sec.~\ref{sec:loop from potential model}. 

\subsection{Modelling the background dynamics}
\label{sec:model with instantaneous transitions}

The action for canonical, single-field models of inflation is 
\begin{equation}
    \label{action single field inflation}
    \mathcal{S} = \int \mathrm{d}^4x\, \sqrt{-g} \; \left[\frac{1}{2} M_p^2 R -\frac{1}{2} (\partial_\mu \phi)^2 -V(\phi) \right] \;,
\end{equation}
where $\phi$ is the inflaton field and $V(\phi)$ its potential. The Hubble rate, $H\equiv \dot a/a$, and the homogeneous inflaton field satisfy the equations
\begin{equation}
\label{backeq}
    3H^2=\frac{\dot{\phi}^2}{2}+V \;, \quad  \ddot{\phi}+3H\dot{\phi}+V_\phi=0 \;,
\end{equation}
and the first and second\footnote{There is an entire series of slow-roll parameters defined by means of derivatives of the Hubble rate, $\epsilon_{i+1} = \dot \epsilon_i /(H \epsilon_i)$, with $i\geq 0$ and $\epsilon_0 = H_\text{in}/H$. In Eq.~\eqref{slow roll parameters} we only report the first two, identifying $\epsilon_1=\epsilon$ and $\epsilon_2 =\eta$.} Hubble slow-roll parameters are defined by 
\begin{equation}
\label{slow roll parameters}
    \epsilon\equiv -\frac{\dot{H}}{H^2} = \frac{1}{2} \frac{\dot \phi^2}{H^2}\;,  \qquad \eta\equiv \frac{\dot{\epsilon}}{H \epsilon} \;. 
\end{equation}
In SR inflation, the inflaton is overdamped $(|\ddot \phi | \ll 3H|\dot \phi|)$ and slowly rolls down its own potential $(\dot \phi ^2 \ll V)$. In this case Eq.~\eqref{backeq} reduces to 
\begin{equation}
    \label{SR background}
    3 H^2 \simeq V\;, \quad 3 H \dot{\phi}+ V_\phi \simeq 0 \;,
\end{equation}
and the slow-roll parameters are small $(\epsilon, \eta \ll 1)$. 

In a phase of USR, the potential becomes sufficiently flat such that the term $V_\phi$ can be neglected in Eq.~\eqref{backeq}, yielding 
\begin{equation}
\label{USR back}
    3 H^2 \simeq V\;, \quad \ddot{\phi}+3H\dot{\phi}\simeq0 \;.
\end{equation}
From this it follows that $\dot \phi \propto a^{-3}$, and therefore, the slow-roll parameters in Eq.~\eqref{slow roll parameters} become
\begin{equation}
\label{slow roll param USR}
    \epsilon \propto a^{-6} \; , \quad \eta \simeq-6 \;. 
\end{equation}
The curvature perturbation in the comoving gauge produced during  inflation, $\zeta_k(\tau)$, can be expressed in terms of the Mukhanov--Sasaki variable, $v_k$, as $\zeta_k\equiv v_k/z$, where $z\equiv a\sqrt{2\epsilon}$. The variable $v_k$ obeys the Mukhanov--Sasaki equation 
\begin{equation}
\label{MS equation}
    \frac{\mathrm{d}^2 v_k}{\mathrm{d}\tau^2}+\Bigg(k^2-\frac{1}{z}\frac{\mathrm{d}^2 z}{\mathrm{d}\tau^2}\Bigg)v_k=0 \;, 
\end{equation}
and the normalisation condition 
\begin{equation}
    \label{normalisation condition}
    v_k'^* v_k - v_k'v_k^* =i \;,
\end{equation}
which follows from quantisation of the curvature perturbation. 

In Ref.~\cite{Kristiano:2022maq},  the authors consider an inflationary scenario featuring SR-USR-SR dynamics where the transitions between different regimes are assumed to be instantaneous.
The large-scale CMB mode exits the horizon during the first SR phase with an almost scale-invariant power spectrum satisfying constraints from CMB measurements~\cite{Planck:2018jri}. After the initial evolution over a region of the potential that supports SR inflation, it is assumed that the inflaton then crosses an approximately flat region triggering a phase of USR. Since the curvature perturbation mode functions are proportional to $\epsilon^{-1/2}$, they become amplified as $\zeta_k \propto a^3$, see Eq.~\eqref{slow roll param USR}. 

The background SR-USR-SR dynamics in Ref.~\cite{Kristiano:2022maq} is modelled using the time evolution of the slow-roll parameters \eqref{slow roll parameters}, i.e. \textit{not} from an explicit potential formulation. This is achieved by connecting together three phases, each one characterised by a constant value of $\eta$: in this case $\eta:0\to-6\to0$, with instant transitions between them. For other examples where this technique is used see, e.g., Refs.~\cite{Byrnes:2018txb,Carrilho:2019oqg,Davies:2021loj}.  A model within this scenario is specified by its values of $\tau_s$ and $\tau_e$, which are defined as the conformal times marking the start and end of the USR phase. The duration of the USR phase, and hence the size of the enhancement of short-scale modes, is determined by the ratio $\tau_e/\tau_s$.

This simplified picture allows extensive analytic computations. For example, the evolution of the scalar perturbation is obtained by matching solutions to the Mukhanov--Sasaki equation~\eqref{MS equation} derived for different constant values of $\eta$. 
Indeed, when $\eta$ is constant, the Fourier modes of the primordial curvature perturbation are given in terms of Hankel functions of the first and second kind as 
\begin{equation} \label{MF}
\zeta_k(\tau)=\frac{-H\tau}{\sqrt{2\epsilon(\tau)}}\left[A_k \sqrt{-\tau} H^{(1)}_{\nu}(-k \tau)+B_k \sqrt{-\tau}H^{(2)}_{\nu}(-k\tau)\right] \;,
\end{equation}
where $\nu$ is defined by
\begin{equation}
    \nu^2=\frac{9}{4}+\frac{3}{2}\eta+\frac{1}{4}\eta^2+\frac{\dot{\eta}}{2H}+\mathcal{O}(\epsilon)  
\end{equation}
and $A_k$ and $B_k$ are $k$-dependent constants. The mode functions and their first derivatives with respect to conformal time are then matched at the transition times as per the Israel junction conditions~\cite{Israel:1966rt,Deruelle_1995}. By applying this procedure, and imposing Bunch-Davies initial conditions for $\zeta_k(\tau)$ during the first SR phase, one can fix the constants $A_k$ and $B_k$. See Ref.~\cite{Kristiano:2022maq} for the explicit form of the mode functions.

Of primary interest when calculating the predictions of an inflationary model, the primordial power spectrum is related to the 2-point correlator of the curvature perturbation in Fourier space according to
\begin{equation}
\label{2-point corr zeta}
   \langle \zeta_{\mathbf{k}}\zeta_{\mathbf{k}'}\rangle=(2\pi)^3\delta(\mathbf{k}+\mathbf{k}')P_\zeta(k)\;,
\end{equation}
where we are implicitly evaluating this quantity at the end of inflation (or as soon as the inflaton evolution becomes adiabatic, i.e. when the curvature perturbation is conserved in the super-horizon regime). From Eq.~\eqref{2-point corr zeta} one can define the dimensionless power spectrum as
\begin{equation}
\label{dimensionless power spectrum}
    \mathcal{P}_{\zeta}(k)=\frac{k^3}{2\pi^2}P_{\zeta}(k)\;.
\end{equation}
The different contributions to $\mathcal{P}_\zeta(k)$ can be organised in a loop-expansion, where the $n$th-loop term contains $n$ unconstrained momentum integrations. In this fashion, at $n$th-loop order we have
\begin{equation}
    \label{loop expansion}
    \mathcal{P}_\zeta(k)=\mathcal{P}_{\zeta}(k)_{ \rm tree}+\mathcal{P}_{\zeta}(k)_{\rm 1-loop}+\cdots +  \mathcal{P}_{\zeta}(k)_{\rm n-loop}\;.
\end{equation}
The first contribution in Eq.~\eqref{loop expansion}, the tree-level scalar power spectrum, 
is calculated by using the linear equation of motion for the curvature perturbation. The purpose of this work is to calculate the 1-loop correction on large-scales due to small-scale modes, enhanced by the transient USR phase. This makes up just one contribution to $\mathcal{P}_{\zeta}(k)_{\rm 1-loop}$. From now on we will refer to it as the 1-loop correction and use $\mathcal{P}_{\zeta}(k)_{\rm 1-loop}$ to identify it.

By employing the analytic solution for $\zeta_k(\tau)$ in Eq.~\eqref{MF}, one can calculate the tree-level dimensionless power spectrum as 
\begin{equation}
    \mathcal{P}_\zeta(k;\tau)_{\rm tree}=\frac{k^3}{2 \pi^2}|\zeta_k(\tau)|^2 \;.
\end{equation}

In Fig.~\ref{fig:PSInstant} we plot the tree-level dimensionless power spectrum at late times ($\tau \rightarrow0$) for two different models with instantaneous transitions described by the above framework. At large scales, the spectra are scale invariant with amplitude $\mathcal{P}_\zeta(k)_\text{tree}= 2.1 \times 10^{-9}$, in accordance with the observed CMB value~\cite{Planck:2018jri}. As we move to shorter scales, the amplitude of the spectra dip down before growing to a large peak. Immediately after the peak, the instantaneous nature of the transitions leads to oscillations in the spectra before they plateau to a constant amplitude. The difference between the two models considered lies in how large the enhancement of the spectrum at peak scales is. The spectrum represented in red is enhanced by 7 orders of magnitude relative to the large scales, so that the peak has amplitude $\mathcal{P}_\zeta(k_\text{peak})_\text{tree}=0.01$. The one plotted in blue is enhanced by 8 orders of magnitude, and we get a peak amplitude $\mathcal{P}_{\zeta}(k_{\text{peak}})_\text{tree}= 0.3$. In this case, the spectrum then plateaus to an amplitude of 0.03. Although the peak amplitude of the second model is too large to be physically realised, we include it in our analysis to highlight the effect of increasing the peak amplitude on the value of the 1LC. Including it also allows us to make further contact with the results stated in Ref.~\cite{Kristiano:2022maq}. We now detail the calculation of the 1-loop correction following the methodology of that work.

\subsection{Calculation of the 1-loop correction}
\label{sec:loop instantaneous model}
In Ref.~\cite{Kristiano:2022maq}, Kristiano \& Yokoyama calculate the 1LC on large scales by employing the In-In formalism, see e.g.~\cite{Adshead:2009cb,Weinberg:2005vy}. Within this framework, one can calculate the expectation value of an operator ${\mathcal{O}}$ at time $\tau$ according to
\begin{equation}
\label{inin}
    \langle {\mathcal{O}}(\tau)\rangle=\bigg \langle \bigg [ \bar{T} \exp \bigg ( i \int^{\tau}_{-\infty(1+i\epsilon)} \mathrm{d} \tau' H_{\rm int}(\tau')\bigg) \bigg ] {\mathcal{O}}(\tau) \bigg [ T \exp \bigg ( -i \int^{\tau}_{-\infty(1-i\epsilon)} \mathrm{d} \tau' H_{\rm int}(\tau')\bigg) \bigg ]\bigg \rangle\;,
\end{equation}
where $H_{\rm int}$ is the interaction Hamiltonian and $\bar{T}$ and $T$ denote the anti-time and time ordering operators respectively. The deformation by $i\epsilon$ of the contour is needed to guarantee that the theory becomes free in the infinite past, i.e. $\ket{\Omega}\rightarrow \ket{0}$ as $\tau \rightarrow -\infty$, where $\ket{\Omega}$ is the vacuum of the interacting theory. We omit this deformation in subsequent sections, since when we go on to compute Eq.~\eqref{inin} in  Sec.~\ref{sec:loop from potential model}, we do so numerically, beginning with a finite lower limit for the time integrals.

To compute the 1LC from Eq.~\eqref{inin}, we set $\mathcal{O}(\tau)=\zeta(\mathbf{p},\tau)\zeta(-\mathbf{p},\tau)$ where $\mathbf{p}$ is the comoving wavenumber of the large scale for which we want to compute the correction. The (dimensionfull) 1LC, $P_\zeta(p;\,\tau)$, is then defined as 
\begin{equation}
    \label{dimensionfull 1LC def}
     \langle \zeta (\mathbf{p},\,\tau) \zeta (\mathbf{-p},\,\tau)\rangle_{\text{1-loop}} = (2\pi)^3 \delta (\mathbf{p-p}) P_\zeta(p;\,\tau)_\text{1-loop}\;,
\end{equation}
from which we obtain the dimensionless 1LC
\begin{equation}
    \label{dimensionless 1LC def}
    \mathcal{P}_\zeta(p;\,\tau)_{\text{1-loop}} = \frac{p^3}{2\pi^2} {P}_\zeta(p;\,\tau)_{\text{1-loop}}  \;.
\end{equation}
\begin{figure}
\centering
  \includegraphics[width=0.4\linewidth]{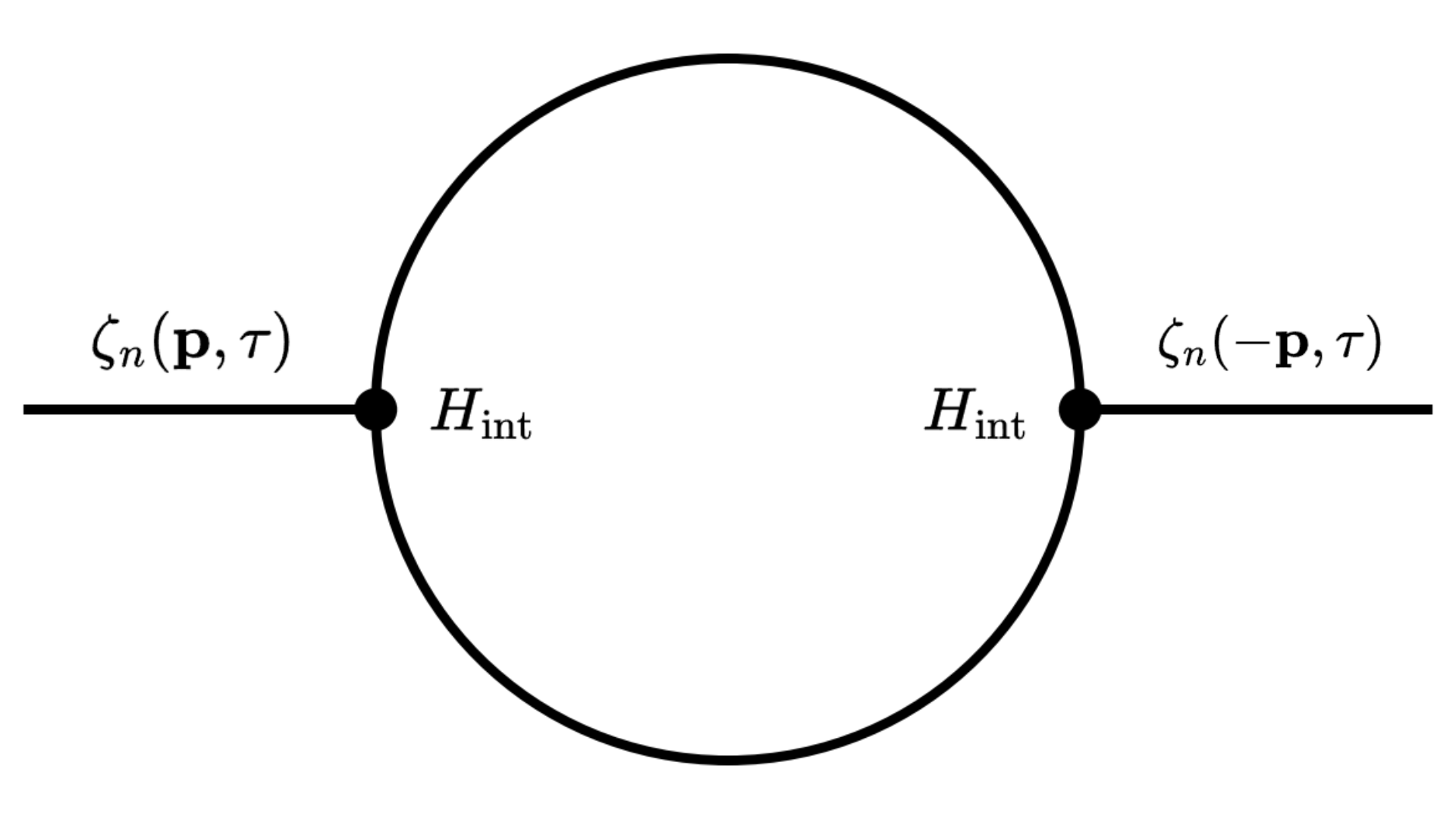}
\caption{One of the diagrams contributing to the 1LC in Eq.~\eqref{dimensionfull 1LC def}. This term arises from cubic interactions with $H_{\rm int}$ given in Eq.~\eqref{int}. In this work, we only consider the contribution to the 1LC from this diagram.}
\label{fig:Diagrams}
\end{figure} 
Since, in this work, we are primarily concerned with reproducing the calculation in Ref.~\cite{Kristiano:2022maq}, but for a potential-driven model, we only consider the contribution to the 1LC from the cubic interaction\footnote{To calculate the total 1LC, one should of course consider the diagram involving quartic interactions for $\zeta$, see e.g. Ref.~\cite{Firouzjahi:2023aum}.} which we represent diagrammatically in Fig.~\ref{fig:Diagrams}. 

By expanding the operators in Eq.~\eqref{inin} containing $H_\text{int}$ up to second order, and taking into consideration time-ordering, one gets
\begin{equation}
        \langle \mathcal{O}(\tau)\rangle =\langle \mathcal{O}(\tau)\rangle^\dag_{(0,2)}+\langle \mathcal{O}(\tau)\rangle_{(1,1)}+\langle \mathcal{O}(\tau)\rangle_{(0,2)} \;,
\end{equation}
where
\begin{align}
         \langle \mathcal{O}(\tau)\rangle_{(1,1)} = \int^\tau_{-\infty}\mathrm{d}\tau_1 \int^\tau_{-\infty}\mathrm{d}\tau_2 \, \langle H_{\rm int} (\tau_1)\hat{\mathcal{O}}(\tau)H_{\rm int}(\tau_2)\rangle  \;,  \\
         \langle \mathcal{O}(\tau)\rangle_{(0,2)} = -\int^\tau_{-\infty}\mathrm{d}\tau_1 \int^{\tau_1}_{-\infty}\mathrm{d}\tau_2 \, \langle \hat{\mathcal{O}}(\tau)H_{\rm int}(\tau_1)H_{\rm int}(\tau_2)\rangle\;.
\end{align}
The cubic-order interaction Hamiltonian to be substituted in the expressions above can be calculated from the cubic action, which is obtained by expanding Eq.~\eqref{action single field inflation} to third order in $\zeta$~\cite{Maldacena_2003, Namjoo_2013}
\begin{equation} 
\label{3rdaction}
\begin{split}
S^{(3)}_{\rm int}&= \mathcal{S}^{(3)}_\text{bulk}+ \mathcal{S}^{(3)}_\text{boundary} \\
& = M_{\rm pl}^2\int \mathrm{d}t \, \mathrm{d}^3x \, \bigg[a^3 \epsilon^2 \zeta\dot{\zeta}^2+a\epsilon^2\zeta(\partial \zeta)^2-2a \epsilon \dot{\zeta}(\partial \zeta)(\partial \chi)\\
&\; +\frac{a^3 \epsilon}{2}\dot{\eta}\zeta^2\dot{\zeta}+\frac{\epsilon}{2a}(\partial\zeta)(\partial\chi)\partial^2 \chi+\frac{\epsilon}{4a}(\partial^2 \zeta)(\partial \chi)^2\\
&\; +2f(\zeta)\frac{\delta L}{\delta \zeta}\bigg\rvert_1 \bigg] + \mathcal{S}^{(3)}_\text{boundary} \;. 
\end{split}
\end{equation}
We distinguish here between the bulk cubic action and the boundary terms, and note that
\begin{equation}
\partial^2 \chi=a^2 \epsilon \dot{\zeta}\, , \,  \frac{\delta L}{\delta \zeta}\bigg\rvert_1=a(\partial^2 \dot{\chi}+H\partial^2 \chi - \epsilon \partial^2 \zeta)
\end{equation}
and
\begin{equation} \label{FOR}
\begin{aligned}
f(\zeta)={}&\frac{\eta}{4}\zeta^2+\frac{1}{H}\zeta\dot{\zeta}\\&+\frac{1}{4 a^2 H^2}[-(\partial \zeta)(\partial \zeta)+\partial^{-2}(\partial_i \partial_j(\partial_i \zeta \partial_j \zeta))]\\&+\frac{1}{2a^2 H}[(\partial \zeta)(\partial \chi)-\partial^{-2}(\partial_i \partial_j(\partial_i \zeta\partial_j\chi))]\;.
\end{aligned}
\end{equation}
The last term in the cubic action \eqref{3rdaction}, proportional to the equations of motion, is usually removed by making a field redefinition of the form~\cite{Maldacena_2003}
\begin{equation}
\label{field redef}
    \zeta\rightarrow \zeta_n +f(\zeta_n)\;.
\end{equation}
Realistic models of inflation end with a SR phase, i.e. the field redefinition is reversed during SR, and therefore the second and fourth terms in Eq.~\eqref{FOR} are suppressed by $\dot{\zeta}$. The third term can also be neglected since it includes spatial derivatives which are suppressed on superhorizon scales, since they include factors of $k/aH\ll 1$ when written in Fourier space. This implies that the field redefinition \eqref{field redef} can be simplified to
\begin{equation}
\label{simplified field redef}
    \zeta\rightarrow \zeta_n +\frac{\eta}{4} \zeta_n^2\;.
\end{equation}

Transforming the cubic action \eqref{3rdaction} by means of the field redefinition \eqref{field redef} is a procedure used extensively in the literature (see e.g.~\cite{Maldacena_2003,Namjoo_2013,Cai:2018dkf,Davies:2021loj}) to calculate the scalar bispectrum, $B_\zeta(k_1,k_2,k_3)$, i.e. the 3-point correlation function of $\zeta$ in Fourier space 
\begin{equation} 
\label{bispec}
    \langle \zeta_{\mathbf{k}_1}\zeta_{\mathbf{k}_2} \zeta_{\mathbf{k}_3}\rangle=(2\pi)^3 \delta(\mathbf{k}_1+\mathbf{k}_2+\mathbf{k}_3) B_\zeta(k_1,k_2,k_3) \, .
\end{equation}
Besides removing the term proportional to the equation of motion, the redefinition \eqref{field redef} leaves the rest of the bulk cubic action in Eq.~\eqref{3rdaction} unchanged, but with $\zeta$ replaced by $\zeta_n$. The action in terms of $\zeta_n$ can then be used to compute the bispectrum of the new field, $B_{\zeta_n}(k_1,k_2,k_3)$. The bispectrum of the original field $\zeta$ is obtained by undoing the field redefinition \eqref{field redef} at the time at which one wishes to evaluate $B_\zeta$, e.g. at the end of inflation. 
Substituting Eq.~\eqref{simplified field redef} into the 3-point correlation function and applying Wick's theorem, we have\footnote{Note that this procedure will also generate terms proportional to $\eta^2$ and $\eta^3$, but these are suppressed by additional factors of $\zeta_n$.} 
\begin{equation}
\label{undo}
\langle \zeta_{\mathbf{k}_1}\zeta_{\mathbf{k}_2} \zeta_{\mathbf{k}_3}\rangle=(2\pi)^3 \delta(\mathbf{k}_1+\mathbf{k}_2+\mathbf{k}_3) \left\{B_{\zeta_n}(k_1,k_2,k_3)+(2\pi)^3\frac{\eta}{4}\left[P_{\zeta_n}(\mathbf{k}_1)P_{\zeta_n}(\mathbf{k}_2)+\textrm{perms.}\right]\right\}\,.
\end{equation}
The first term of Eq.~\eqref{undo} is the bispectrum of the new field, $\zeta_n$, while the second term is due to the procedure of undoing the field redefinition \eqref{simplified field redef}. Note that the latter is proportional to $\eta$ evaluated at late times. This implies that, for a model ending in a phase of SR inflation, the second contribution in Eq.~\eqref{undo}, proportional to $\eta$, will be slow-roll suppressed, and so the bispectrum of the original and redefined fields are equivalent up to SR corrections.

One could follow a similar line of thought to assess the impact of the field redefinition on the calculation of the 1LC. After performing the field redefinition \eqref{field redef}, which removes the last term in $\mathcal{S}^{(3)}_\text{bulk}$, the leading interaction term in presence of SR-USR-SR dynamics is the one proportional to $\dot \eta$. This is because the remaining interactions in Eq.~\eqref{3rdaction} are slow-roll suppressed, while $\dot{\eta}$ becomes transiently large during the SR-USR and USR-SR transitions. For this reason, in previous works 
(e.g.~\cite{Kristiano:2022maq,Tasinato:2023ukp,Bhattacharya:2023ysp,Choudhury:2023vuj}), the dominant term from the cubic action to be used in the In-In master formula \eqref{inin} is assumed to be the three-point interaction
\begin{equation}
\label{int}
H_{\rm int}=-\frac{1}{2}M_{\rm pl}^2\int \; \mathrm{d}^3 x \; \epsilon \eta'a^2 \zeta_n'\zeta_n^2 \;. 
\end{equation}
Furthermore, any potentially relevant term in $\mathcal{S}^{(3)}_{\rm boundary}$ is also taken care of by performing the field redefinition~\cite{Arroja_2011,Burrage:2011hd}. This allows us to calculate the 1LC in analogy with the bispectrum calculation reviewed above. The 1LC will be given by the 1LC of the new field, $\zeta_n$, and a contribution from undoing the field redefinition. 
Since at the end of inflation the contribution from undoing the field redefinition is negligible, one might expect that the two 1LCs coincide~\cite{Kristiano:2023scm}, just as the two bispectra in Eq.~\eqref{undo} do. In the case of the 1LC, however, it has been argued by Firouzjahi in Ref.~\cite{Firouzjahi:2023bkt} that the non-linear relationship between $\zeta$ and $\zeta_n$ induces quartic interactions for $\zeta_n$, which need to be taken into account. 
In this work, since our principal aim is to understand how the analysis of Ref.~\cite{Kristiano:2022maq} differs when one works from a potential-driven model, we proceed to calculate the contribution only from cubic interactions, as done in Ref.~\cite{Kristiano:2022maq}. 
The form and size of contributions from induced quartic-order interactions would be an interesting subject for future work.

We also note that the 1LC can be calculated without employing a field redefinition. In Ref.~\cite{Fumagalli:2023hpa}, Fumagalli explored the possibility of working directly from the action in terms of $\zeta$ in Eq.~\eqref{3rdaction}. In this approach, one would need to take into account relevant terms in both the bulk and boundary actions. Fumagalli originally claimed that this method yielded a volume-suppressed result, $\text{1LC}\propto (p/k)^3$, i.e. the 1LC is suppressed by the separation of scales between the large-scale mode, $p$, and small-scale modes, $k$, where $p\ll k$. However, this calculation was recently revisited by Firouzjahi in Ref.~\cite{Firouzjahi:2023bkt}, who finds a 1LC for models with instantaneous transitions of a similar size to that claimed by Kristiano \& Yokoyama.

It is beyond the purpose of this work to resolve discrepancies between the various expressions for the 1LC in the literature generated by different methodologies. 
As underlined above, our primary objective is instead to work within the same framework as Ref.~\cite{Kristiano:2022maq} and verify whether the results for the 1LC calculated analytically for a model with instant transitions in Ref.~\cite{Kristiano:2022maq} are unchanged when calculated numerically from an explicit potential formulation. 

By performing the field redefinition \eqref{field redef} and adopting Eq.~\eqref{int} as the interaction Hamiltonian,
the authors of Ref.~\cite{Kristiano:2022maq} find
\begin{equation}
\label{1LC1}
    \mathcal{P}_\zeta(p;\,\tau)_\text{1-loop} =\frac{1}{4}\epsilon^2(\tau_e)a^4(\tau_e)\Delta\eta(\tau_e)^2 |\zeta_p(\tau)|^2 \times 16 \int \frac{\mathrm{d}^3 k}{(2\pi)^3}\bigg[ \frac{p^3}{2\pi^2}|\zeta_k|^2\Im(\zeta_p'\zeta_p^*)\Im(\zeta_q'\zeta_q^*)\bigg]_{\tau=\tau_e} \;, 
\end{equation}
where the 1LC is evaluated at a late time $\tau$ (e.g. at the end of inflation), $p$ is the large-scale mode, $q=|\mathbf{k}-\mathbf{p}|$ and the integral is performed over all modes $k$. Also recall that $\tau_e$ is the conformal time signalling the end of USR, and so $\Delta\eta(\tau_e)$ is the change in $\eta$ at $\tau_e$. In this case $\Delta\eta(\tau_e)=6$. 

We note that Eq.~\eqref{1LC1} captures the contribution to the 1LC only from the USR-SR transition. The contribution from the SR-USR transition is neglected; since none of the modes have been enhanced by the time this transition occurs, it is expected to be subdominant. We test in Sec.~\ref{sec:robust} the validity of this assumption within our numerical study. 

Evaluating Eq.~\eqref{1LC1} requires careful considerations. First of all, the integral in Eq.~\eqref{1LC1} extends over arbitrarily small scales, i.e. high wavenumber $k$, yielding a UV-divergent result. This means that a proper computation of the 1LC cannot be carried out without regularisation and renormalisation of Eq.~\eqref{1LC1}. 
However, following Ref.~\cite{Kristiano:2022maq}, one could instead restrict the integration range over peak scales only. 
We define $k_s$ and $k_e$ to be the scales that crossed the horizon at the beginning and end of the USR phase respectively. 
Integrating between these scales returns a finite result from Eq.~\eqref{1LC1}. 
Consideration in Ref.~\cite{Kristiano:2022maq} of the UV divergence, and how to address it using an order by order renormalisation scheme, results in the same bound on the amplitude of $\mathcal{P}_\zeta(k_\text{peak})_\text{tree}$ as that obtained from the 1LC calculation restricted to peak scales. 
This is in line with the expectation that the majority of the contribution to the 1LC should come from peak scales, as these scales are greatly enhanced compared to any other scale.
While we believe further investigation of this issue is important, we reiterate that the purpose of this work is to apply the same methods as in Ref.~\cite{Kristiano:2022maq} to a model derived from a potential and compare the results. Henceforth, we assume that the finite contribution obtained by evaluating Eq.~\eqref{1LC1} over peak scales yields a reasonable order-of-magnitude approximation to the 1LC.
\begin{figure}
  \centering
  \includegraphics[width=0.6\linewidth]{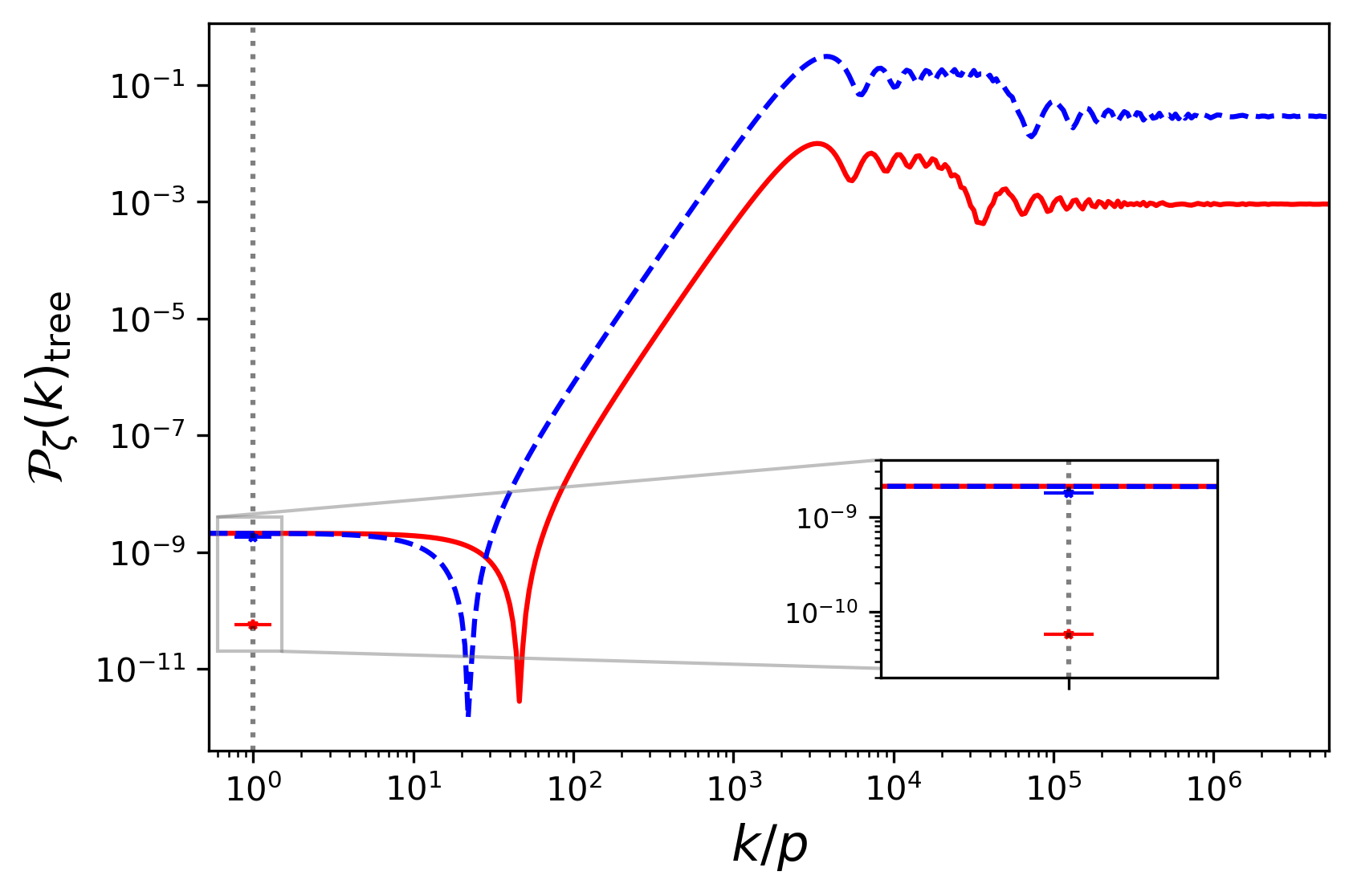}
\caption{Tree-level dimensionless scalar power spectra for two instantaneous models with different peak-amplitude enhancements. The dashed-blue spectrum has a peak amplitude $\mathcal{P}_\zeta(k_{\rm peak})_{\rm tree}=0.3$, whereas the solid-red spectrum has a peak amplitude $\mathcal{P}_\zeta(k_{\rm peak})_{\rm tree}=0.01$. A dotted, vertical line marks the position of the scale $p$. The 1LCs to each model at a large scale $p$ are displayed using blue and red markers.}
\label{fig:PSInstant}
\end{figure}

Proceeding with Eq.~\eqref{1LC1}, and restricting the integration to peak scales $k_s \leq k\leq k_e$, we can obtain a value for the 1LC at a large scale $p$ for the two instantaneous models in Fig.~\ref{fig:PSInstant}. The two models are specified by selecting values for the conformal times $\tau_s$ and $\tau_e$, or equivalently $k_s$ and $k_e$ (since $k_s=-1/\tau_s$ and $ k_e=-1/\tau_e$), see Tab.~\ref{tab:parameters}. 
The 1LCs for each model are displayed in Fig.~\ref{fig:PSInstant} using red and blue markers. 
We find 
that the solid-red model, with a peak amplitude $\mathcal{P}_\zeta(k_{\rm peak})_{\rm tree}=0.01$ yields a loop-correction on large scales at late times $\mathcal{P}_{\zeta}(p)_{\rm 1-loop}=5.77 \times 10^{-11}$. The dashed-blue model, meanwhile, has a peak amplitude $\mathcal{P}_\zeta(k_{\rm peak})_{\rm tree}=0.3$ and a loop-correction $\mathcal{P}_{\zeta}(p)_{\rm 1-loop}=1.79 \times 10^{-9}$.

The first thing to notice is that the size of the 1LC computed from Eq.~\eqref{1LC1} depends sensitively on the peak amplitude of the dimensionless power spectrum. 
The results in Fig.~\ref{fig:PSInstant} demonstrate that models producing a power spectrum peak of $\mathcal{O}(10^{-1})$ encounter significant issues with perturbativity.  
Indeed, the 1LC for the dashed-blue model in Fig.~\ref{fig:PSInstant}, is the same size as the amplitude measured by CMB experiments, $\mathcal{P}_\zeta(k_{\rm CMB})=2.1 \times 10^{-9}$~\cite{Planck:2018jri}.
In this case, perturbativity at CMB scales cannot be trusted.
This finding may not be cause for too much concern, though, since such large peak amplitudes 
are likely ruled out already due to overproduction of PBHs~\cite{Gow:2020bzo}. 

Meanwhile, if we calculate the 1LC for the solid-red model in Fig.~\ref{fig:PSInstant}, which has a realistic peak amplitude for PBH production, we find that it is only $3\%$ of the observed amplitude at CMB scales. This is consistent with 
perturbativity 
on large-scales.

It is important to connect these findings with the results stated 
in Ref.~\cite{Kristiano:2022maq}. 
If we were to employ the analytic formula presented in their Eq.~(35)\footnote{We enumerate this equation using the second version of Ref.~\cite{Kristiano:2022maq}.} to calculate the 1LC for the solid-red model displayed in Fig.~\ref{fig:PSInstant}, associating the peak of the spectrum with $\Delta^2_{s \,(\rm{PBH})}$ in that formula, we would find a value of $3.85\times10^{-10}$ for the 1LC -- significantly larger than the one we find. 
The reason for this is that the analytical analysis of Ref.~\cite{Kristiano:2022maq} neglects the transition back to slow-roll.
Due to this transition, $\mathcal{P}_\zeta(k)_\text{tree}$ relaxes back to a plateau value which is smaller than the amplitude at the peak by a factor of $4$, see Fig.~\ref{fig:PSInstant}. 
For this reason, their formula would be extremely accurate if $\Delta^2_{s \,{\rm (PBH)}}$ is instead associated
with the \emph{final} plateau value the spectrum acquires at $k>k_\text{peak}$.  
With this change, Eq.~(35) in Ref.~\cite{Kristiano:2022maq} yields a value for the 1LC of $3.51\times10^{-11}$, now in much better agreement with our numerical result. 
Clearly, an important consequence of this discussion is that the bound quoted in Ref.~\cite{Kristiano:2022maq} on the peak amplitude of the power spectrum is weakened.
In fact, this bound is relaxed even further because the analytical approach of Ref.~\cite{Kristiano:2022maq} also does not take into account the oscillations in $\mathcal{P}_\zeta(k)_\text{tree}$ at peak scales. 
Since PBH formation is exponentially sensitive to $\mathcal{P}_\zeta(k_\text{peak})$, it is really 
the amplitude of the first oscillation seen in Fig.~\ref{fig:PSInstant} that should enter into any bound. 
To summarise, while Eq.~(35) of Ref.~\cite{Kristiano:2022maq} might lead one to believe that the 1LC could be of the same order of magnitude as the tree-level term for peaks relevant for PBH production, we have argued that their conclusion is too strong, and shown with our numerical analysis that for a peak of ${\cal O} (10^{-2})$ the 1LC on large scales is around two orders of magnitude smaller than the linear contribution.
A similar conclusion was drawn also from the analysis in Ref.~\cite{Franciolini:2023lgy}.
With the dashed-blue model displayed in Fig.~\ref{fig:PSInstant} we show an example of a spectrum that yields a 1-loop correction comparable to the tree-level amplitude.
From this, we conclude that for models with instant transitions one should have $\mathcal{P}_\zeta(k_\text{peak})\ll 0.3$ in order to avoid perturbativity concerns\footnote{We note that this is in perfect agreement with the bound quoted in Eq.~(37) in the second version of Ref.~\cite{Kristiano:2022maq}, so long as $\Delta^2_{s \,{\rm (PBH)}}$ is properly interpreted as the final plateau value of the spectrum.}.  

One last point to consider is that we have been calculating the 1LC to a large scale, $p$, which in reality is much shorter than the CMB scale, $k_\text{CMB}=0.05\,\text{Mpc}^{-1}$; see Fig.~\ref{fig:PSInstant}.
We do not expect the value of the 1LC to depend on the selected large-scale $p$, as long as $p$ is chosen in the range of scales corresponding to the plateau in $\mathcal{P}_\zeta(k)_\text{tree}$. 
In this case, Eq.~\eqref{1LC1} is independent of $p$. 
Indeed, $|\zeta_q|^2 \approx |\zeta_k|^2$, since $p\ll k$, and the two terms  
${p^3}|\zeta_p|^2/({2\pi^2})$ and $\Im(\zeta'_p\zeta_p^*)$ are in practice independent of $p$.  
The former is just the scale-invariant\footnote{Kristiano \& Yokoyama also include the effect of the tilt in the derivation of their bound in Eq.~(37) of Ref.~\cite{Kristiano:2022maq}. Indeed, while the toy model with instantaneous transitions they employ yields a scale-invariant power spectrum on large scales, CMB measurements constrain it to be slightly red-tilted~\cite{Planck:2018jri}. One might think that by including this effect the resulting 1LC would be larger, as for fixed $\mathcal{P}_\zeta(k_\text{peak})$ a larger enhancement would be required. Nevertheless, we have checked that including the effect of a red tilt doesn't affect the size of the 1LC by more than 1\%. 
}  tree-level power spectrum on large scales, and
the latter is scale-invariant, as demonstrated in Ref.~\cite{Kristiano:2022maq}. 
In conclusion, the 1LC calculated in an instantaneous model for  
the intermediate large scale $p$ is a good proxy for the value of the
1LC at the CMB scale.

While these results are derived analytically for models with instantaneous transitions, we now proceed to consider the 1LC calculated numerically from an analytic potential. 
Smoothing off the instantaneous nature of transitions might reduce the size of the 1LC even further~\cite{Riotto:2023hoz,Franciolini:2023lgy,Firouzjahi:2023ahg}. 

\section{1-loop correction on large scales: SR-USR-SR dynamics from an explicit potential formulation}
\label{sec:loop from potential model} 
A first numerical calculation of the 1LC for a model with non-instantaneous transitions appeared in Ref.~\cite{Franciolini:2023lgy}. In that work, the evolution of the background, i.e. $\eta(N)$, was modelled analytically with a function that allows for smoothed transitions. In this sense, our work goes a step further as we derive the 1LC directly from an explicit potential. 

In Sec.~\ref{sec:potentialmod}, we introduce the inflationary potential used to mimic the behaviour of the SR-USR-SR instantaneous model studied in Ref.~\cite{Kristiano:2022maq}, and outlined in the previous section. In Sec.~\ref{sec: the in in calculation} we generalise the In-In formalism calculation above to derive an analytic expression for the 1LC which does not assume that transitions are instantaneous and is suitable for numerical computation. The results from this potential model and the instantaneous one are then compared in Sec.~\ref{sec:compare}.

\subsection{Potential model}
\label{sec:potentialmod}

\begin{figure}
\centering
\includegraphics[width=0.6\linewidth]{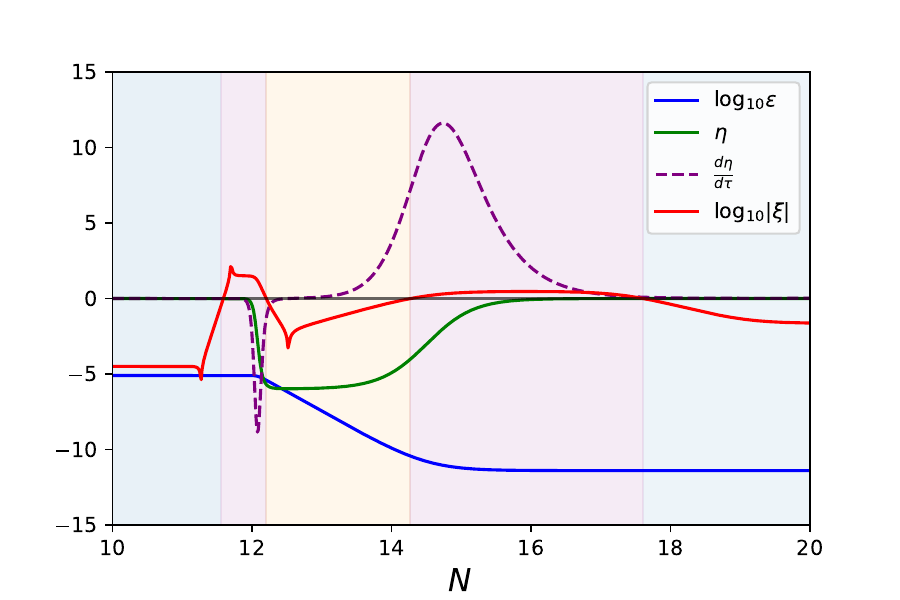}
\caption{Background evolution of $\epsilon$, $\eta$, $\eta'$ and $|\xi|$ for the `Matched Peak' potential model plotted in green on the left of Fig.~\ref{fig:PS}. Phases of constant $\eta$ (shaded blue regions denote SR, shaded orange regions denote USR) and transitions (shaded purple regions) in the model are defined by the size of $|\xi|$, see Eq.~\eqref{Transition} and the related discussion.}
\label{fig:back}
\end{figure}

We work with the potential
\begin{equation}
\label{potential}
    V(\phi)=V_0 \left[p_0+p_1\left(\log\left(\cosh\left(p_2 \phi\right)\right)+\left(p_2+p_3\right)\phi\right)\right] \;,
\end{equation}
which produces an inflationary dynamics similar to that of a model with instant transitions, but now all background quantities vary smoothly \footnote{Following Ref.~\cite{Cai:2018dkf}, one could define the character of the transitions by means of the parameter $h\equiv -6\sqrt{\epsilon_V/\epsilon}$, defined at the end of USR. In this work we are primarily interested in (possible) changes in the value of the 1LC due to smooth background evolution. In other words, the comparison we make is between the models with instantaneous transitions studied in Ref.~\cite{Kristiano:2022maq}, characterised by $h=-6$, and a potential model, yielding $h\to0$. We leave the study of models featuring highly sharp transitions $(h\ll -6)$ for future work.}; see Fig.~\ref{fig:back}.
This potential can be thought of as a smoothed version of Starobinsky's linear piecewise model~\cite{Starobinsky:1992ts}, relatives of which have been used to study the effect of USR on PBH formation and non-Gaussianity, see e.g. Refs.~\cite{Carrilho:2019oqg, Pi_2023,domènech2023exact}.
\begin{table}[]
\resizebox{\columnwidth}{!}{%
\begin{tabular}{l|lllll|ll|}
\cline{2-8}
                                      & \multicolumn{5}{l|}{Potential Model}                                                                                                                & \multicolumn{2}{l|}{Instantaneous Model} \\ \cline{2-8}
      & \multicolumn{1}{l|}{$V_0$}               & \multicolumn{1}{l|}{$p_0$} & \multicolumn{1}{l|}{$p_1$}             & \multicolumn{1}{l|}{$p_2$} & $p_3$ & \multicolumn{1}{l|}{$k_s/p$}    & $k_e/k_s$    \\ \hline
\multicolumn{1}{|l|}{Matched Peak}    & \multicolumn{1}{l|}{$4 \times 10^{-12}$} & \multicolumn{1}{l|}{1}     & \multicolumn{1}{l|}{$5\times 10^{-7}$} & \multicolumn{1}{l|}{$4\times 10^3$}  & 5.5   & \multicolumn{1}{l|}{$1.06\times 10^3$}      & \multicolumn{1}{l|}{8.69}     \\ \hline
\multicolumn{1}{|l|}{Matched Plateau} & \multicolumn{1}{l|}{$4 \times 10^{-12}$} & \multicolumn{1}{l|}{1}     & \multicolumn{1}{l|}{$5\times10^{-7}$}  & \multicolumn{1}{l|}{$4 \times 10^3$}  & 2     & \multicolumn{1}{l|}{$1.22 \times 10^3$}     & \multicolumn{1}{l|}{15.5}    \\ \hline
\end{tabular}%
}
\caption{Parameter values for the instantaneous and potential models used to produce the tree-level dimensionless power spectra, $\mathcal{P}_\zeta(k)_\text{tree}$, in Fig.~\ref{fig:PS}.}
\label{tab:parameters}
\end{table}
We set the values of the parameters in Eq.~\eqref{potential} in such a way that the resulting tree-level dimensionless power spectra can be meaningfully compared to those produced by the instantaneous models discussed in Sec.~\ref{sec:loop with instantaneous transitions}. 
We select two sets of parameters, whose values are listed in Tab.~\ref{tab:parameters}, and represent in Fig.~\ref{fig:PS} with green lines the corresponding  numerical results for $\mathcal{P}_\zeta(k)_\text{tree}$.
In particular, for the spectrum on the right hand-side of Fig.~\ref{fig:PS} the potential parameters are fixed so that the amplitude of the plateau at short-scales matches that of the instantaneous model (the dashed-blue model from Fig.~\ref{fig:PSInstant}). 
In this case, the potential model can be thought of as a proper `smoothing' of the corresponding instantaneous model and behaves in a similar way to the smoothed $\eta$ models already studied in Ref.~\cite{Franciolini:2023lgy} (see Sec.~\ref{sec:compare} for further details). 
Alternatively, on the left hand-side of Fig.~\ref{fig:PS} we fix the potential parameters so that the peak amplitude of the power spectrum from the potential matches the peak amplitude of the power spectrum from the instantaneous model (the solid-red model from Fig.~\ref{fig:PSInstant}). 
In this case, the comparison is motivated by physical considerations, since PBH production depends exponentially on the precise value of $\mathcal{P}_\zeta(k_\text{peak})$.
 
We numerically solve the background equations~\eqref{backeq} using the initial conditions $\{\dot{\phi}(0)=0, \, \phi(0)=\phi_0\}$, where the value $\phi_0$ determines the number of e-folds of evolution before the transition to USR. 
In Fig.~\ref{fig:back}, we plot the evolution of relevant background quantities. 
The evolution of the slow-roll parameters $\epsilon$ (solid-blue) and $\eta$ (solid-green) demonstrate how similarly this potential model behaves to the instantaneous SR-USR-SR one. 
The value of $\epsilon$ is constant during the SR periods, and decays during USR as predicted by Eq.~\eqref{slow roll param USR}. 
The second slow-roll parameter, $\eta$, varies between 0 and -6 (SR and USR respectively), but now does so smoothly. 

In the presence of smooth evolution one needs to set a criterion to distinguish between the SR and USR phases, and the transitions between them. 
In instantaneous models, the transitions between phases of SR and USR occur at precise, well-defined times, i.e. a phase of constant $\eta$ is unambiguously defined by $\eta'=0$. 
In a potential-driven model, $\eta'$ (dashed-purple line in Fig.~\ref{fig:back}) varies smoothly and is never zero for a continuous length of time, posing the issue of defining the transition periods.
For the purpose of this work, it is nevertheless important to quantitatively identify these periods. 
Indeed, it is during the transitions that $\eta'$ grows large and the interaction in Eq.~\eqref{int} dominates, yielding a potentially large contribution to the 1LC. 
To define the transition periods, we employ the third Hubble slow-roll parameter, 
\begin{equation}
    \xi\equiv\frac{1}{\eta}\frac{\mathrm{d}\eta}{\mathrm{d}N} \;,
\end{equation}
whose evolution is displayed with a solid-red line in Fig.~\ref{fig:back}.  
We define the transition periods according to the criterion
\begin{equation}
\label{Transition}
|\xi| \geq 1 \,.
\end{equation}
The vertical boundaries between coloured regions in all figures are determined by this condition,  
and shaded purple regions correspond to the transition periods themselves.

The curvature perturbation mode functions, $\zeta_k$, evolve according to the Mukhanov-Sasaki equation \eqref{MS equation}, which we solve numerically by imposing Bunch-Davies initial conditions. For their real and imaginary parts of $\zeta_k$ these read
\begin{subequations}
\begin{align}
\label{MFinitial 1}
        \Re(\zeta_k(\tau_i))=\frac{1}{\sqrt{2k}}\cos{\left(k \tau_i\right)}\;, \quad
        \Re(\zeta'_k(\tau_i))=-\frac{1}{a H}\sqrt{\frac{k}{2}}\sin{\left(k \tau_i\right)}\;, \\
\label{MFinitial 2}
        \Im(\zeta_k(\tau_i))=-\frac{1}{\sqrt{2k}}\sin{\left(k \tau_i\right)}\;, \quad 
        \Im(\zeta'_k(\tau_i))=-\frac{1}{a H}\sqrt{\frac{k}{2}}\cos{\left(k \tau_i\right)}\;.
\end{align}
\end{subequations}
In the expressions above, $\tau_i$ is a time well before the mode in question crossed the horizon. 
In our numerical computations the initial conditions \eqref{MFinitial 1}--\eqref{MFinitial 2} are usually imposed for each mode function 5 e-folds before the time of horizon crossing.

\subsection{Preparing the In-In analytical expression for numerical evaluation}
\label{sec: the in in calculation}
We evaluate the 1LC for the potential model within the same procedure, and under the same assumptions, as in Ref.~\cite{Kristiano:2022maq}. In other words, we apply the In-In formalism, perform a field redefiniton and assume the cubic interaction~\eqref{int} is the dominant one. For more details see Sec.~\ref{sec:loop instantaneous model}. In order to get an analytical expression for the 1LC suitable for efficient numerical evaluation, we find it useful to take a step back and review part of the analytical In-In calculation. 

By applying Wick's theorem and retaining only connected diagrams, one finds
\begin{equation}
\label{Full1LC}
\begin{split}
    P_\zeta&(p;\,\tau)_\text{1-loop} = \frac{1}{4}\int_{-\infty}^{\tau} \mathrm{d}\tau_1\; \epsilon(\tau_1) \eta'(\tau_1) a^2(\tau_1)   \int_{-\infty}^{\tau} \mathrm{d}\tau_2  \; \epsilon(\tau_2) \eta'(\tau_2) a^2(\tau_2) \int \frac{\mathrm{d}^3k}{(2\pi)^3}\;\times \\
    & \zeta^*_p(\tau)\zeta_p(\tau) \Big[ 4 \zeta'_p(\tau_1) \zeta_k(\tau_1) \zeta_q(\tau_1) \zeta^{\prime *}_p(\tau_2) \zeta_k^* (\tau_2)  \zeta_q^* (\tau_2)  + 8 \zeta_p(\tau_1) \zeta'_k(\tau_1)  \zeta_q(\tau_1) \zeta_p^*(\tau_2) \zeta_k^* (\tau_2) \zeta^{\prime *}_q (\tau_2) \\
    & + 8 \zeta_p(\tau_1) \zeta'_k(\tau_1) \zeta_q(\tau_1)  \zeta_p^*(\tau_2) \zeta^{\prime *}_k (\tau_2)  \zeta_q^* (\tau_2) + 8 \zeta'_p(\tau_1) \zeta_k(\tau_1) \zeta_q(\tau_1) \zeta_p^*(\tau_2) \zeta_k^* (\tau_2) \zeta^{\prime *}_q (\tau_2)  \\
    & + 8 \zeta_p(\tau_1) \zeta'_k(\tau_1) \zeta_q(\tau_1)  \zeta^{\prime *}_p (\tau_2)  \zeta_k^* (\tau_2) \zeta_q^*(\tau_2) \Big]  \\
    & - \frac{1}{2}\int_{-\infty}^{\tau} \mathrm{d}\tau_1\; \epsilon(\tau_1) \eta'(\tau_1) a^2(\tau_1)   \int_{-\infty}^{\tau_1} \mathrm{d}\tau_2  \; \epsilon(\tau_2) \eta'(\tau_2) a^2(\tau_2) \int \frac{\mathrm{d}^3k}{(2\pi)^3}\;\times \\
    & \zeta_p(\tau)^2\;  \Re\Big[ 4   \zeta^{\prime *}_p (\tau_1) \zeta_k(\tau_1) \zeta_q(\tau_1) \zeta^{\prime *}_p (\tau_2) \zeta_k^*(\tau_2) \zeta_q^* (\tau_2) + 8 \zeta_p^* (\tau_1) \zeta_k(\tau_1) \zeta'_q(\tau_1) \zeta_p^* (\tau_2) \zeta_k^*(\tau_2) \zeta^{\prime *}_q (\tau_2)   \\
    & + 8 \zeta_p^* (\tau_1)  \zeta_k(\tau_1)  \zeta'_q(\tau_1) \zeta_p^* (\tau_2) \zeta^{\prime *}_k(\tau_2) \zeta_q^* (\tau_2)  + 8 \zeta_p^* (\tau_1)  \zeta_k(\tau_1) \zeta'_q(\tau_1)\zeta^{\prime *}_p (\tau_2)  \zeta_k^*(\tau_2) \zeta_q^* (\tau_2)  \\
    & + 8 \zeta^{\prime *}_p (\tau_1) \zeta_k(\tau_1) \zeta_q(\tau_1) \zeta_p^* (\tau_2) \zeta^{\prime *}_k(\tau_2) \zeta_q^* (\tau_2) \Big]  \;.
\end{split}
\end{equation} 
Up to this point, no assumption has been made on the quality of the SR-USR or USR-SR transitions, i.e. this expression is equally valid for instantaneous and potential models. 

For the instantaneous models discussed in Sec.~\ref{sec:loop with instantaneous transitions},
the only contributions to the 1LC come from the times when $\eta'\neq 0$, i.e. at 
$\tau=\tau_s$ and $\tau=\tau_e$ when the transitions occur.
In this case, the two time integrations in Eq.~\eqref{Full1LC} can be performed analytically.
In Ref.~\cite{Kristiano:2022maq}, the authors neglect the contribution from the initial SR-USR transition, since at that time the mode functions are not yet enhanced, and obtain Eq.~\eqref{1LC1}.
On the other hand, when dealing with a potential model, $\eta'$ is never truly zero and the time integrals receive contributions from a range of times. 

Eq.~\eqref{Full1LC} is given by the sum of two terms, which we distinguish by their differing domains for the time integrals. 
We have checked that, upon fixing finite domains for the time and momentum integrals (see the discussion below Eq.~\eqref{analytic1LC1}), 
numerically computing the 1LC using the expression given in Eq.~\eqref{Full1LC} is difficult. 
This is due to 
delicate cancellations between results from
the two separate terms. A better approach consists in simplifying Eq.~\eqref{Full1LC} analytically before performing the numerical evaluation (a similar method for simplifying this expression was outlined in Ref.~\cite{Franciolini:2023lgy}). 
First, by noting that the measure of the momentum integral, performed over all $\mathbf{k}$ space, is invariant under translations one can group the five contributions within each of the two terms in Eq.~\eqref{Full1LC} into a single one 
\begin{equation}
\label{1LCs1}
\begin{split}
    P_\zeta(p;\,\tau)_\text{1-loop} =& \frac{1}{4}\int_{-\infty}^{\tau} \mathrm{d}\tau_1\; \epsilon(\tau_1) \eta'(\tau_1) a^2(\tau_1)   \int_{-\infty}^{\tau} \mathrm{d}\tau_2  \; \epsilon(\tau_2) \eta'(\tau_2) a^2(\tau_2) \int \frac{\mathrm{d}^3k}{(2\pi)^3}\;\times \\
    &\quad\Big[4  \frac{\mathrm{d}}{\mathrm{d}\tau_1}\big[\zeta_p(\tau_1) \zeta_k (\tau_1) \zeta_q(\tau_1)\big] \zeta^*_p (\tau)\zeta_p(\tau) \frac{\mathrm{d}}{\mathrm{d}\tau_2}\big[\zeta_p^*(\tau_2) \zeta_k^* (\tau_2) \zeta_q^*(\tau_2)\big]\Big]  \\
     &-\frac{1}{2}\int_{-\infty}^{\tau} \mathrm{d}\tau_1\; \epsilon(\tau_1) \eta'(\tau_1) a^2(\tau_1)   \int_{-\infty}^{\tau_1} \mathrm{d}\tau_2  \; \epsilon(\tau_2) \eta'(\tau_2) a^2(\tau_2) \int \frac{\mathrm{d}^3k}{(2\pi)^3}\;\times \\
    & \quad\Re\Big[4  \frac{\mathrm{d}}{\mathrm{d}\tau_1}\big[\zeta^*_p(\tau_1) \zeta_k (\tau_1) \zeta_q(\tau_1)\big] \zeta_p (\tau)\zeta_p(\tau) \frac{\mathrm{d}}{\mathrm{d}\tau_2}\big[\zeta_p^*(\tau_2) \zeta_k^* (\tau_2) \zeta_q^*(\tau_2)\big]\Big]  \;.
\end{split}
\end{equation} 
We then notice that the time integral in the first term is performed over an entire (infinite) square in the $(\tau_1,\,\tau_2)$ space, while in the second term the domain is restricted to the lower-half triangle. 
Moreover, the first-term integrand is symmetric under the transformation $\{\tau_1 \to \tau_2\;, \tau_2\to\tau_1\}$, since the integrand must be real and therefore equal to its complex conjugate.   
This property implies that the result from integrating over the entire (infinite) square is simply given by two times the result obtained from integrating over the lower-half triangle.  
From these considerations, the two terms in Eq.~\eqref{1LCs1} can be combined under the same integral sign, 
\begin{equation}
\label{1LC intermediate}
\begin{split}
    P_\zeta(p;\,\tau)_\text{1-loop} = \frac{1}{2}&\int_{-\infty}^{\tau} \mathrm{d}\tau_1\; \epsilon(\tau_1) \eta'(\tau_1) a^2(\tau_1)   \int_{-\infty}^{\tau_1} \mathrm{d}\tau_2  \; \epsilon(\tau_2) \eta'(\tau_2) a^2(\tau_2) \int \frac{\mathrm{d}^3k}{(2\pi)^3}\;\times \\
    &\Big\{4  \frac{\mathrm{d}}{\mathrm{d}\tau_1}\big[\zeta_p(\tau_1) \zeta_k (\tau_1) \zeta_q(\tau_1)\big] \zeta^*_p (\tau)\zeta_p(\tau) \frac{\mathrm{d}}{\mathrm{d}\tau_2}\big[\zeta_p^*(\tau_2) \zeta_k^* (\tau_2) \zeta_q^*(\tau_2)\big]\\
     &-\Re\Big[4  \frac{\mathrm{d}}{\mathrm{d}\tau_1}\big[\zeta^*_p(\tau_1) \zeta_k (\tau_1) \zeta_q(\tau_1)\big] \zeta_p (\tau)\zeta_p(\tau) \frac{\mathrm{d}}{\mathrm{d}\tau_2}\big[\zeta_p^*(\tau_2) \zeta_k^* (\tau_2) \zeta_q^*(\tau_2)\big] \Big]\Big\}  \;.
\end{split}
\end{equation}
Now that the two terms are under the same integral sign, it is easier to simplify equal contributions from the two integrands.
By expanding $\zeta_p$ in terms of its imaginary and real parts, $\zeta_p=\zeta_p^R + i \zeta_p^I$, one can rewrite $\zeta_p^*(\tau_1)=\zeta_p(\tau_1)-2i\zeta^I_p(\tau_1)$ and $\zeta_p (\tau)=\zeta^*_p(\tau)+2i\zeta^I_p(\tau)$. By substituting these into the second integrand in Eq.~\eqref{1LC intermediate}, one of the resulting terms cancels with the first integrand, while the remaining contributions can be written as  
\begin{equation}
\begin{split}
    P_\zeta(p;\,\tau)_\text{1-loop} = \frac{1}{2}&\int_{-\infty}^{\tau} \mathrm{d}\tau_1\; \epsilon(\tau_1) \eta'(\tau_1) a^2(\tau_1)   \int_{-\infty}^{\tau_1} \mathrm{d}\tau_2  \; \epsilon(\tau_2) \eta'(\tau_2) a^2(\tau_2) \int \frac{\mathrm{d}^3k}{(2\pi)^3}\;\times \\
     \Im\Big[&8  \zeta_p(\tau) \frac{\mathrm{d}}{\mathrm{d}\tau_2}\big[\zeta^*_p(\tau_2) \zeta^*_k (\tau_2) \zeta^*_q(\tau_2)\big] \times \\ 
     &\Big\{\zeta_p^I(\tau)\frac{\mathrm{d}}{\mathrm{d}\tau_1}\big[\zeta_p^*(\tau_1) \zeta_k (\tau_1) \zeta_q(\tau_1)\big]-\zeta^*_p(\tau)\frac{\mathrm{d}}{\mathrm{d}\tau_1}\big[\zeta_p^I(\tau_1) \zeta_k (\tau_1) \zeta_q(\tau_1)\big]\Big\}\Big]  \;.
\end{split}
\end{equation}
This expression can be simplified further by expanding $\zeta^*_p(\tau_1)=\zeta_p^R(\tau_1)-i \zeta_p^I(\tau_1)$ and $\zeta^*_p(\tau)=\zeta_p^R(\tau)-i \zeta_p^I(\tau)$, which leads to
\begin{equation}
\label{analytic1LC1}
\begin{split}
    P_\zeta(p;\,\tau)_\text{1-loop} = \frac{1}{2}&\int_{-\infty}^{\tau} \mathrm{d}\tau_1\; \epsilon(\tau_1) \eta'(\tau_1) a^2(\tau_1)   \int_{-\infty}^{\tau_1} \mathrm{d}\tau_2  \; \epsilon(\tau_2) \eta'(\tau_2) a^2(\tau_2) \int \frac{\mathrm{d}^3k}{(2\pi)^3}\;\times \\
     \Im\Big[&8  \zeta_p(\tau) \frac{\mathrm{d}}{\mathrm{d}\tau_2}\big[\zeta^*_p(\tau_2) \zeta^*_k (\tau_2) \zeta^*_q(\tau_2)\big]\times\\
     &\bigg\{\zeta_p^I(\tau)\frac{\mathrm{d}}{\mathrm{d}\tau_1}\big[\zeta^R_p(\tau_1) \zeta_k (\tau_1) \zeta_q(\tau_1)\big]-\zeta^R_p(\tau)\frac{\mathrm{d}}{\mathrm{d}\tau_1}\big[\zeta_p^I(\tau_1) \zeta_k (\tau_1) \zeta_q(\tau_1)\big]\bigg\}\Big]  \;.
\end{split}
\end{equation}
This is the analytical expression that we find most efficient for numerical evaluation. We stress that no assumption has been made about the nature of the transition to derive Eq.~\eqref{analytic1LC1}.

When numerically evaluating the background and perturbations evolution produced by a potential model, one usually measures time in e-foldings of expansion, defined by $N\equiv \int_{t_i}^{t_f}   H(t) \mathrm{d}t$. 
We therefore transform all derivatives with respect to cosmic time in Eq.~\eqref{analytic1LC1} into derivatives with respect to e-folds. Moreover, due to the large separation of scales between a large-scale mode that crossed the horizon during SR, $p$ in this case, and the USR-enhanced modes $k$, we can simplify Eq.~\eqref{analytic1LC1} by making the replacement $q=|\mathbf{k-p}|\approx k$. 

In analogy with Ref.~\cite{Kristiano:2022maq} (see also Ref.~\cite{Franciolini:2023lgy}), we restrict the momentum integral in Eq.~\eqref{analytic1LC1} over the finite range of peak scales. 
While in the instantaneous model identifying $k_s$ and $k_e$ is unambiguous, in realistic models one has to identify a criterion to distinguish the SR and USR phases, and the transitions between them. 
As discussed in Sec.~\ref{sec:potentialmod}, we utilize the criterion \eqref{Transition}, which in turn defines unambiguously the modes crossing the horizon during the USR phase. 
In passing, we note that this  range of modes 
is quite close to the range of scales integrated over in the instantaneous models. 
This ensures that the 1LC results calculated for the instantaneous model and those obtained from the potential one can be compared fairly, e.g. differences between them cannot be accounted for by different momentum domains.   

Continuing to follow the methodology of Ref.~\cite{Kristiano:2022maq}, we have assumed that the dominant cubic interaction is the one proportional to $\eta'$, see Eq.~\eqref{int}. 
We therefore have to integrate Eq.~\eqref{analytic1LC1} over all times during which $\eta'$ is large, so that we're sure that all relevant contributions to the 1LC have been included. 
For this reason, we choose to restrict the domain of the nested time integrals in Eq.~\eqref{analytic1LC1} from a time, $N_i$, around an e-fold before the beginning of the first transition up until a time, $N_f$, around an e-fold after the end of the second transition. 
We note that this choice allows us to compare the magnitude of the contribution to the 1LC from the first and second transitions, and therefore to test the assumption taken in Ref.~\cite{Kristiano:2022maq}, where the first transition, at $\tau=\tau_s$, is neglected.  
In Sec.~\ref{sec:robust} we verify that extending either of these domain limits has no significant effect on the resulting 1LC. 

By implementing in Eq.~\eqref{analytic1LC1} the considerations discussed above, we obtain an expression for the dimensionless 1LC which we can evaluate numerically
\begin{equation}
\label{1LC2}
\begin{split}
    \mathcal{P}_\zeta(p;\, N)&_{\text{1-loop}} = \frac{p^3}{\pi^4}\int_{N_i}^{N_f} \mathrm{d}N_1\; \epsilon(N_1) \frac{\mathrm{d}\eta(N_1)}{\mathrm{d} N_1}a^2(N_1)   \int_{N_i}^{N_1} \mathrm{d}N_2  \; \epsilon(N_2) \frac{\mathrm{d}\eta(N_2)}{\mathrm{d} N_2} a^2(N_2) \\
    & \int_{k_s}^{k_e} \mathrm{d}k \;k^2 \, \Im\Big[  \zeta_p(N) a(N_1) H(N_1)a(N_2)H(N_2) \frac{\mathrm{d}}{\mathrm{d}N_2}\big[\zeta^*_p(N_2) \zeta^*_k (N_2)^2\big] \\
    &\quad\quad \quad\quad\quad\times \Big\{\zeta_p^I(N)\frac{\mathrm{d}}{\mathrm{d}N_1}\big[\zeta_p^R(N_1) \zeta_k (N_1)^2 \big]-\zeta^R_p(N)\frac{\mathrm{d}}{\mathrm{d}N_1}\big[\zeta_p^I(N_1) \zeta_k (N_1)^2\big]\Big\}\Big]\;. 
\end{split}
\end{equation}
We emphasise again here that the large scale, $p$, for which we compute the 1LC is typically much smaller than a realistic CMB scale, $k_\text{CMB}=0.05\,\text{Mpc}^{-1}$, i.e. $p>k_\text{CMB}$. This is due to numerical complications: it is not possible to track with sufficient precision the evolution of the mode function $\zeta_k(N)$ and its velocity for long times\footnote{See, e.g., the rapid decay of $\zeta'$ during SR in Fig.~\ref{fig:MF}.}, e.g. for the $\sim 55$ e-folds of observable inflation
corresponding to the horizon crossing of a realistic CMB mode~\cite{Liddle:2000cg}. 
We choose $p$ such that it is sufficiently separated from the peak scales, and it is located on the plateau of $\mathcal{P}_\zeta(k)_\text{tree}$ before the dip. 
We verify explicitly in Sec.~\ref{sec:robust} that decreasing
the size of the comoving wavenumber $p$ in Eq.~\eqref{1LC2} yields no significant change in the numerically evaluated 1LC.

We evaluate Eq.~\eqref{1LC2} by using the \textsc{vegas} Monte-Carlo integration package~\cite{Lepage:2020tgj}, and discuss the results in the following section.

\subsection{Instantaneous \textit{vs} potential models: comparing results for the 1-loop correction}
\label{sec:compare}
\begin{figure}
\centering
\begin{subfigure}{.49\textwidth}
  \centering
  \includegraphics[width=\linewidth]{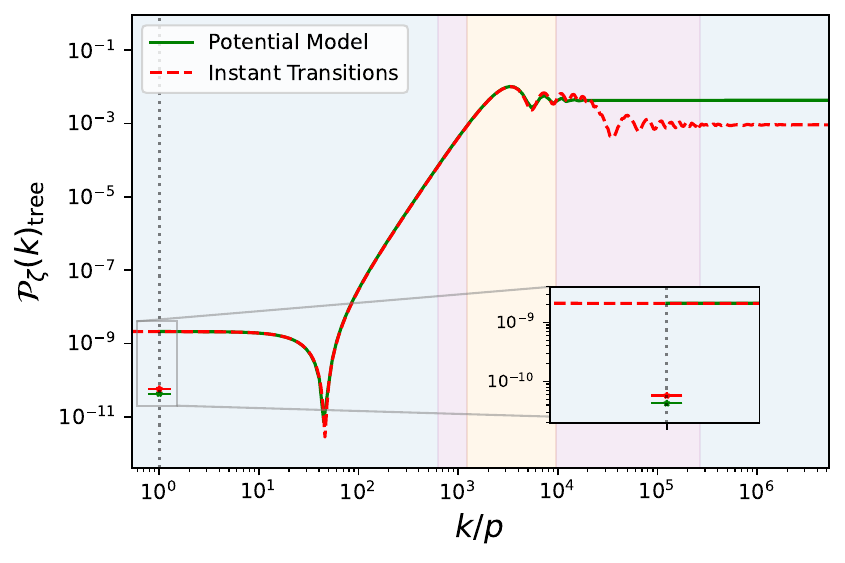}
\end{subfigure}
\begin{subfigure}{.49\textwidth}
  \centering
  \includegraphics[width=\linewidth]{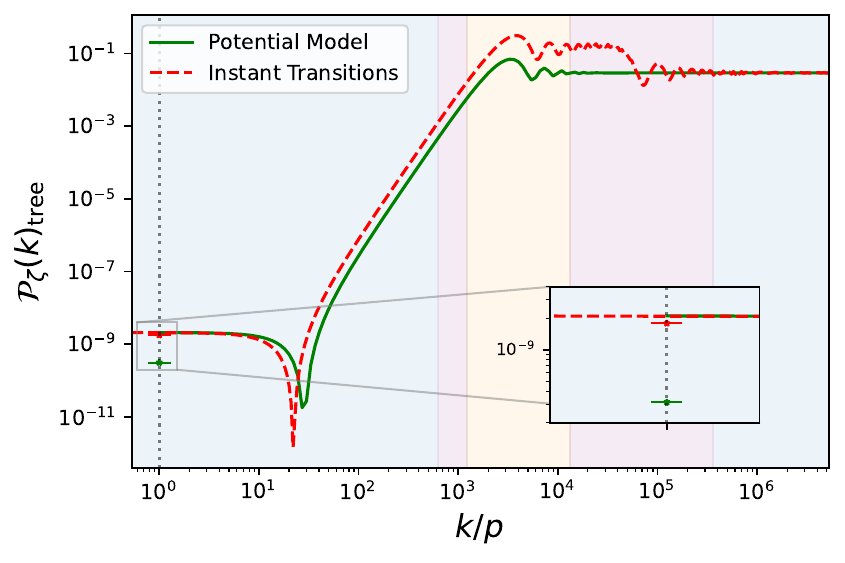}
\end{subfigure}
\caption{Tree-level dimensionless scalar power spectra for a potential model (solid-green line) and a model with instantaneous transitions (dashed-red line) and their 1LCs at the large scale $p$. Two cases are considered: matched peak amplitude (left) and matched plateau amplitude (right). The instantaneous spectrum on the left (right) is the same as the solid-red (dashed-blue) spectrum in Fig.~\ref{fig:PSInstant}. The red and green markers display the size of the 1LC to each spectrum at the large scale $p$. 
A dotted, vertical line marks the position of the scale $p$. Shading indicates the different phases of evolution for the potential model as defined via Eq.~\eqref{Transition}. Comoving scales in blue (purple) regions exit the horizon during SR (the SR-USR transition period and vice versa). Comoving scales in orange regions exit during USR. The USR modes are those over which the integrals in Eq.~\eqref{1LC1} and Eq.~\eqref{1LC2} are performed.}
\label{fig:PS}
\end{figure}
In Fig.~\ref{fig:PS} green markers display the value of the 1LC to the dimensionless power spectrum at a large scale, $p$, for the two models with potential \eqref{potential} and parameters listed in Tab.~\ref{tab:parameters}. 
These are calculated by numerically evaluating Eq.~\eqref{1LC2}. 
In the same figure, we also show in red the results obtained for the analytical models with instantaneous transitions discussed in Sec.~\ref{sec:loop instantaneous model}. These are calculated using Eq.~\eqref{1LC1}, restricting the momentum integral to range over the same scales $k_s$ and $k_e$ defined in the potential model. Note that we work with the same instantaneous models previously shown in Fig.~\ref{fig:PSInstant}.

In the left panel of Fig.~\ref{fig:PS}, which presents the spectra for an instantaneous and a potential model with the same value for $\mathcal{P}_\zeta(k_\text{peak})_\text{tree}$, we see that the 1LC calculated in the two cases have a similar magnitude. We do, however, observe a slightly smaller result for the potential model: 
the size of the 1LC is $5.77\times10^{-11}$ for the instantaneous model 
and $4.26\pm0.02\times10^{-11}$ for the potential one\footnote{The errors here correspond to the standard deviation ($1\sigma$) of the weighted average of the result of the Monte-Carlo numerical integration.}.

For the models displayed in the right panel of Fig.~\ref{fig:PS}, whose tree-level power spectra plateau to the same value, we find instead $3.14\pm0.017 \times10^{-10}$ for the potential model and $1.79\times10^{-9}$ for the instantaneous model.
In other words, the 1LC for the instantaneous model is significantly larger than for the potential model when we match the amplitude of $\mathcal{P}_\zeta(k)_\text{tree}$ at the second SR plateau.  

\begin{figure}
\centering
\begin{subfigure}{.45\textwidth}
  \centering
  \includegraphics[width=\linewidth]{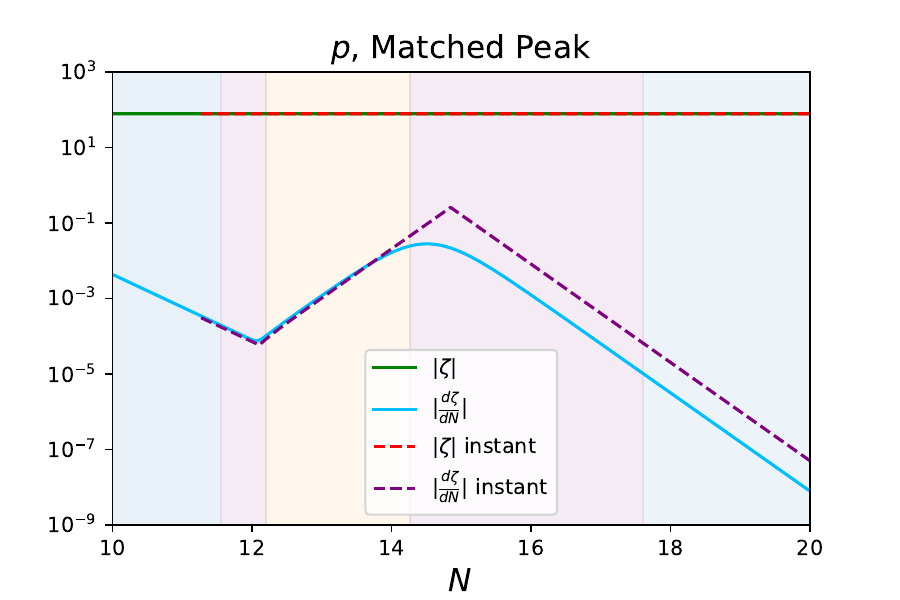}
\end{subfigure}
\begin{subfigure}{.45\textwidth}
  \centering
  \includegraphics[width=\linewidth]{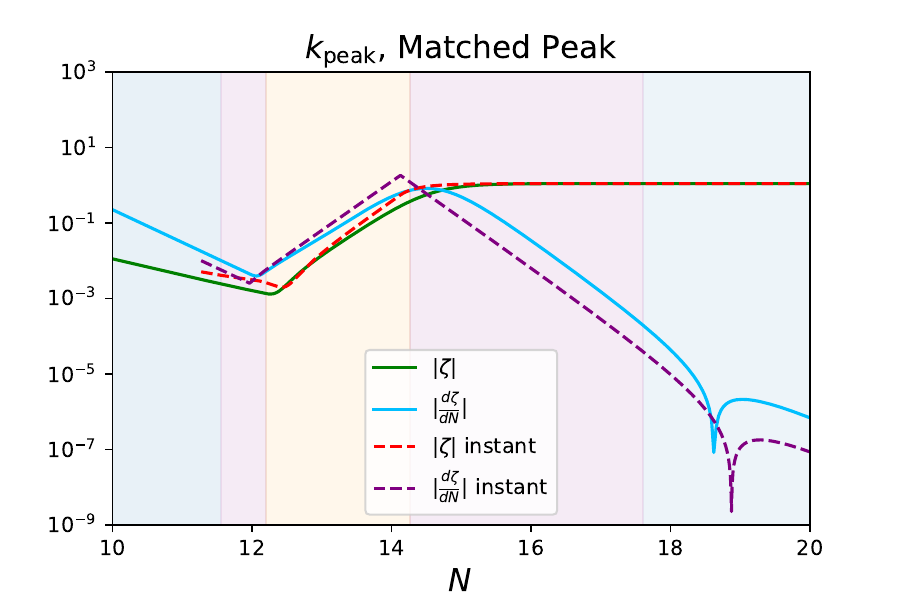}
\end{subfigure}
\centering
\begin{subfigure}{.45\textwidth}
  \centering
  \includegraphics[width=\linewidth]{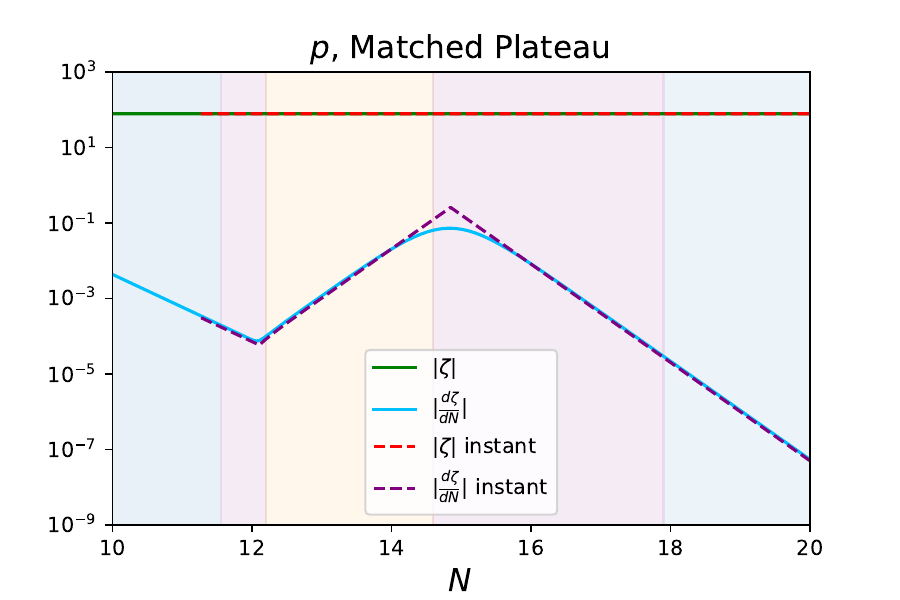}
\end{subfigure}
\begin{subfigure}{.45\textwidth}
  \centering
  \includegraphics[width=\linewidth]{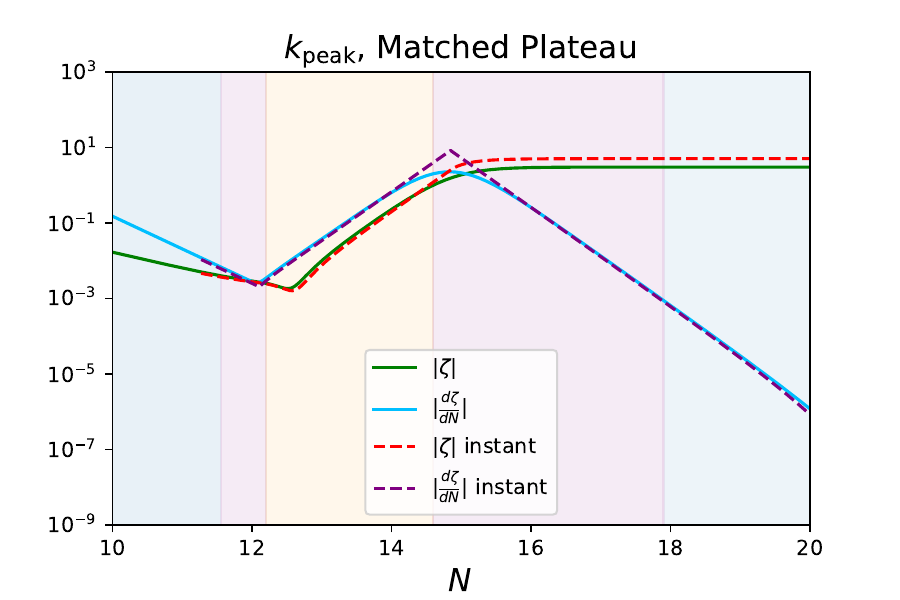}
\end{subfigure}
\caption{Evolution of the magnitude of the curvature mode function, $\zeta$, and its derivative with respect to e-folds, $\mathrm{d}\zeta/\mathrm{d}N$, for a large-scale mode, $p$, (left panels) and the peak-scale mode, $k_{\rm peak}$ (right panels). In each panel we compare the results obtained for the instantaneous (dashed lines) and potential (solid lines) models. The top (bottom) row shows the evolution for the models with matched peak amplitude (matched plateau amplitude), see also Fig.~\ref{fig:PS}.}
\label{fig:MF}
\end{figure}

The results represented in Fig.~\ref{fig:PS} can be explained by inspecting the evolution of the curvature mode function and its derivative, which we represent in Fig.~\ref{fig:MF}. 
One can get an intuition on why this is the case by inspecting Eq.~\eqref{1LC1} (even thought it is derived for a model with instantaneous transitions). Indeed, Eq.~\eqref{1LC1} shows that the size of the 1LC is in part determined by the magnitude of $\zeta$ and $\zeta'$.

In Fig.~\ref{fig:MF}, $\zeta_p$ crossed the horizon long before the onset of the USR phase (in our model it crosses at $N=5$), therefore its amplitude is practically unaffected by it. 
On the other hand, $\zeta_{k_\text{peak}}$ grows during the USR phase and evolves subsequently to a constant value during the transition back to SR. 
We note here that the final amplitude of $\zeta_{k_\text{peak}}$ is different between the instantaneous and potential models when we match the amplitude of the plateaus. 
In particular, $\zeta_{k_\text{peak}}$ is smaller in the potential model. 
We will see the consequence of this in the following. 

As concerns the evolution of the derivative of $\zeta$, we see in all panels that it decays during the first SR period (${\rm{d}\zeta}/{\rm{d}N}\propto a^{-2}$), then grows during USR ($\propto a^{3}$) and decays again afterwards ($\propto a^{-3}$) as the system transitions out of USR. This decay relaxes to the expected rate ($\propto a^{-2}$) for SR at times later than those displayed in the figure. Overall, this results in a large `bump' in the mode function derivative, which is partly responsible for the size of the 1LC.

Inspecting the bottom-row plots in Fig.~\ref{fig:MF}, which display the mode function evolution for the models with matched plateau amplitude (see the right panel of Fig.~\ref{fig:PS}), we see that the derivative of $\zeta$ in the instantaneous model forms a sharp `envelope' for the corresponding quantity calculated from the potential model. 
The derivatives are identical for most of the evolution, except around the transition back to SR, when the results from the potential model smooth the sharp peak obtained in the model with instantaneous transitions. 
This characteristic evolution of $\mathrm{d}\zeta/\mathrm{d}N$ justifies why the potential model is interpreted as a smoothed version of the instantaneous one when the plateaus are matched.  
As anticipated above, smoothing out the instantaneous transition back to SR typically leads to a power spectrum with reduced peak amplitude. 
Similar considerations have already been made in Ref.~\cite{Franciolini:2023lgy}, where instant/smooth transitions are modelled directly from the evolution of $\eta(N)$.
In this case, smoothing the evolution of $\eta(N)$, results in a shorter (integrated) duration of USR (see also~\cite{Cole:2022xqc}), which yields a smaller amplification of the power spectrum on short scales. 
The rounding-off of the local peak in $\mathrm{d}\zeta/\mathrm{d}N$, as well as the reduced magnitude of $\zeta$ on peak scales, explain why the 1LC calculated from a potential model is smaller than the one obtained assuming instant transitions. We find a reduction of almost one order of magnitude, see Fig.~\ref{fig:PS}.  
Since our primary interest is in models which can lead to PBH production, it is important to additionally compare potential and instantaneous models that yield a tree-level power spectra with the same peak amplitude. This is because the process of PBH production depends exponentially on $\mathcal{P}_\zeta(k_\text{peak})$. 
We represent the mode function evolution for the two models, whose spectra are displayed in the left panel of Fig.~\ref{fig:PS}, in the top-row panels of Fig.~\ref{fig:MF}. 
We note that while the final amplitude of $\zeta_{k_\text{peak}}$ is, by construction, the same in both models, 
the peak amplitude of the derivatives are slightly larger in the instantaneous model than in the potential model. 
This explains why we obtain a value for the 1LC $26\%$ smaller in the latter case.

These findings suggest that for models with fixed peak amplitude, $\mathcal{P}_\zeta(k_\text{peak})$,
the size of the 1LC is only minimally changed in a model with a smoothed, potential-driven transition from USR to SR compared with an instantaneous one. 
For a model with a peak amplitude of 0.01, suitably large for PBH production, the 1LC is $3\%$ of the size of the tree-level amplitude of the power spectrum at large scales when calculated for an instantaneous model, and $26\%$ smaller than this when working with a potential-driven model instead.
In other words, simply working with a potential rather than an instantaneous model would not appear to solve the problem of the large magnitude of the 1LC originally found in Ref.~\cite{Kristiano:2022maq}.

\section{Robustness of the results}
\label{sec:robust}
In this section we investigate further the numerical results presented in Sec.~\ref{sec:compare}. 
In particular, we identify the periods in the nested time integral in Eq.~\eqref{1LC2} that contribute the most to the final result (Sec.~\ref{sec: the shape of the integrand}) and test the robustness of the 1LC against changes in the upper and lower limit of the time integral (Sec.~\ref{sec: change limits of time integration}). Additionally, we investigate the effect of increasing the range of the momentum integration (Sec.~\ref{sec: change the upper scale limit}). Finally, while in Sec.~\ref{sec:compare} we calculate the 1LC for a large/small scale separation $p/k_\text{peak}=10^{-4}$, in Sec.~\ref{sec: dependence of 1LC on p} we verify that our results are independent of shifting $p$ to larger and larger scales. 

\subsection{The shape of the integrand}
\label{sec: the shape of the integrand}
Highlighting its dependence on the integration variables $k$, $N_1$ and $N_2$, we rewrite here the integrand in Eq.~\eqref{1LC2},

\begin{equation}
\begin{split}
\label{Integrand}
    \mathcal{I}(k,N_1,N_2)&= \frac{p^3 k^2}{\pi^4} \epsilon(N_1) \frac{\mathrm{d}\eta(N_1)}{\mathrm{d} N_1}a^2(N_1) \epsilon(N_2) \frac{\mathrm{d}\eta(N_2)}{\mathrm{d} N_2} a^2(N_2) \\
   &\times \Im\Big[  \zeta_p(N) a(N_1) H(N_1)a(N_2)H(N_2) \frac{\mathrm{d}}{\mathrm{d}N_2}\big[\zeta^*_p(N_2) \zeta^*_k (N_2)^2\big] \\
    &\qquad \times \Big\{\zeta_p^I(N)\frac{\mathrm{d}}{\mathrm{d}N_1}\big[\zeta_p^R(N_1) \zeta_k (N_1)^2 \big]-\zeta^R_p(N)\frac{\mathrm{d}}{\mathrm{d}N_1}\big[\zeta_p^I(N_1) \zeta_k (N_1)^2\big]\Big\}\Big]\;.
\end{split}
\end{equation}

\begin{figure}
\includegraphics[width=\linewidth]{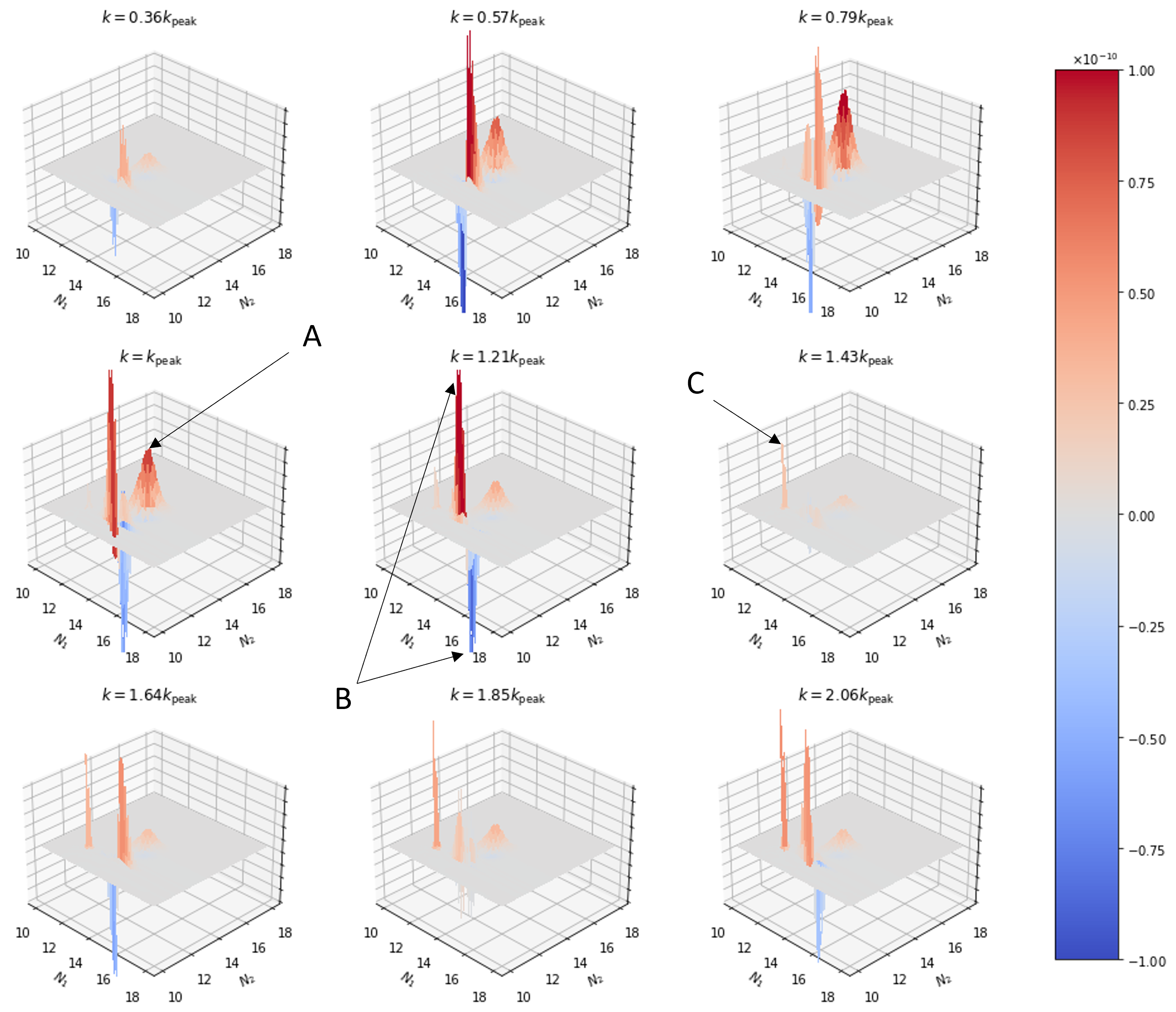}
\caption{ \label{fig:Integrand} Shape of the integrand $\mathcal{I}(k,N_1,N_2)$ for fixed comoving (small-scale) wavenumber $k$. The integrand is calculated for the `Matched Peak' potential model on the left of Fig.~\ref{fig:PS}. In each plot $\mathcal{I}(k,N_1,N_2)$ is displayed on a linear scale over the $(N_1,N_2)$-plane. Note that in Eq.~\eqref{1LC2} we only integrate over the $N_1\geq N_2$ region, therefore in each panel the integrand is set to zero everywhere outside this. Three distinct features of the integrand are labelled `A', `B' and `C', and are discussed in Sec.~\ref{sec: the shape of the integrand}.}
\end{figure}
In Fig.~\ref{fig:Integrand} we represent $\mathcal{I}(k,N_1,N_2)$ as a function of $N_1$ and $N_2$ for $9$ $k$-modes ranging over the peak scales from $k_s$ to $k_e$. The model used to generate these plots is the `Matched Peak' one, see Tab.~\ref{tab:parameters} for the model parameters, and the left panel of Fig.~\ref{fig:PS} for its spectrum.

\begin{figure}
\centering
\begin{subfigure}{.58\textwidth}
  \centering
  \includegraphics[width=\linewidth]{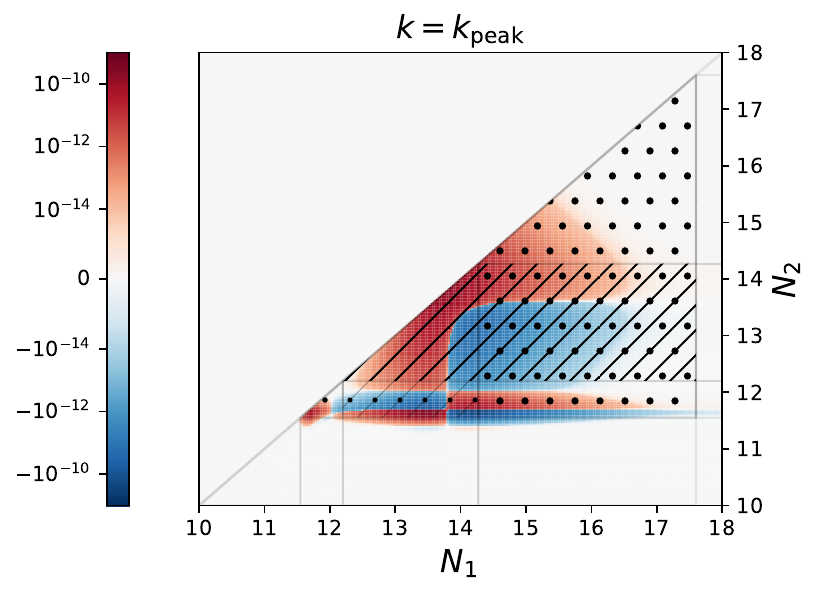}
\end{subfigure}
\begin{subfigure}{.41\textwidth}
  \centering
  \includegraphics[width=\linewidth]{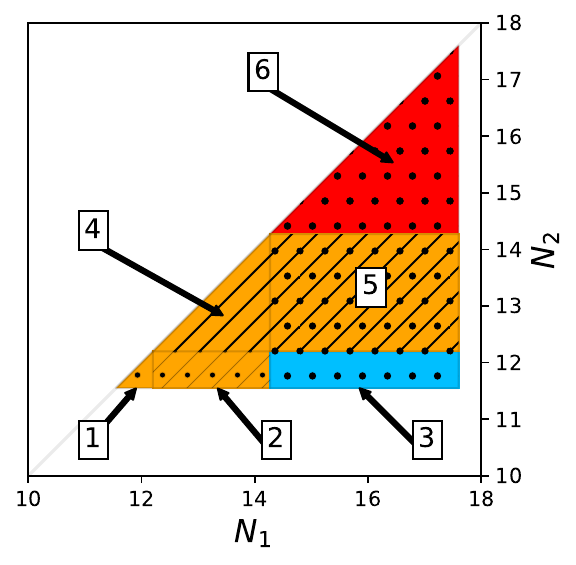}
\end{subfigure}
\caption{\label{fig:IntCol} \textit{Left:} The value of the integrand $\mathcal{I}(k,N_1,N_2)$, represented on a logarithmic colour scale as a function of $N_1$ and $N_2$ for fixed $k=k_{\rm peak}$. The integrand is calculated for the `Matched Peak' potential model on the left of Fig.~\ref{fig:PS}. The relevant areas of the  $(N_1,N_2)$-plane have been hatched according to the dynamics of the model. Solid-line hatches indicate that for $N_a \in \{N_1,N_2\}$, $\{N_a\,:\,12.2\leq N_a\leq14.3\}$, i.e. at least one of the time variables takes a value corresponding to the system being in a phase of USR inflation. Dotted hatching, on the other hand, indicates that for $N_a \in \{N_1,N_2\}$, $\{N_a \, : \, 11.6\leq N_a\leq12.2\} \cup \{N_a \, : \, 14.3\leq N_a\leq17.6\}$, i.e. at least one of the time variables takes a value corresponding to the system transitioning between SR and USR (or vice versa). Regions with no hatching indicate that one of the time variables takes a value during which the system is in SR. \textit{Right:} Regions in the $(N_1,N_2)$-plane contributing to the nested time integral in Eq.~\eqref{1LC2}. We color each region according to their contribution to the final 1LC value after the momentum integral in Eq.~\eqref{1LC2} over all scales between $k_s$ and $k_e$ has been performed. 
The area shaded in blue contributes less than $1\%$ of the total 1LC, regions shaded in orange contribute between $1\%$ and $10\%$, and the one shaded red contributes more than $10\%$. We label each of these coloured regions with a number from 1 to 6 and list in Tab.~\ref{tab:intregion} each of their contributions to the 1LC. Uncoloured (white) regions correspond to areas of the $(N_1,N_2)$-plane that are either not integrated over in Eq.~\eqref{1LC2} (since $N_2>N_1$), or are integrated over, but contribute less than $0.1\%$ of the total 1LC.}
\end{figure}
In Fig.~\ref{fig:IntCol} we plot the same, but for just the peak scale, $k_{\rm peak}$, and display the data as a logarithmic colourmap. In Fig.~\ref{fig:Integrand} we observe, and label, three dominant features in the integrand:
\renewcommand{\labelenumi}{\Alph{enumi})}
\begin{enumerate}
    \item A large contribution coming from times at which $N_1 \approx N_2$ and both are times during either the USR-SR transition ($14.3<N_a\leq17.6)$ or during the USR phase ($12.2<N_a\leq14.3$). This contribution is largest at, and close to, the peak scale $k_{\rm peak}$. Further from the peak, where the spectrum is oscillating, the size of this contribution also oscillates.
    \item Two large contributions, but with opposite signs, when $N_1$ is a time during either the USR phase ($12.2<N_1\leq14.3$) or the USR-SR transition ($14.3<N_1\leq17.6$) and $N_2$ is a time during the SR-USR transition ($11.6<N_2\leq12.2$).
    \item A contribution that grows with the value of $k$, becoming visible just after $k=k_{\rm peak}$, which comes from times at which $N_1 \approx N_2$ and both are times during the SR-USR transition ($11.6<N_a\leq12.2$).
\end{enumerate}
The first listed feature corresponds, partly, to the contribution of the USR-SR transition to the 1LC. This is expected to yield the bulk of the 1LC since this is when both $\eta'$ is largest and the peak-scale mode functions have been enhanced. 

We can verify that the USR-SR transition provides the majority of the contribution to the 1LC by modifying the integration limits of Eq.~\eqref{1LC2} such that we only integrate over times at which the USR-SR transition is occurring. In fact, we can understand where all the different contributions to the total 1LC are coming from by repeating this approach: split the $(N_1,N_2)$-plane into different regions and then integrate Eq.~\eqref{1LC2} over each of them in isolation. 
We show this in Fig.~\ref{fig:IntCol}. In the left panel, we plot the value of the integrand on a logarithmic colour scale and divide the $(N_1,N_2)$-plane in different regions according to the dynamics of the model at the times $N_1$ and $N_2$. In the right panel, we identify six particularly relevant regions and colour them according to what percentage of the total 1LC they contribute when integrated over.  
\begin{table}[]
\resizebox{\columnwidth}{!}{%
\begin{tabular}{|l|l|l|l|}
\hline
Region 
& $N_1$ &$N_2$
& $\times 10^{-11}$, contribution to the 1LC
\\ \hline
1      
& SR-USR &SR-USR       
& 0.42           
\\ 
\hline
2      
& SR-USR &USR            
& 0.1            
\\ 
\hline
3      
& SR-USR &USR-SR 
& -0.06
\\ 
\hline
4      
& USR & USR     & 0.36            \\ 
\hline
5      
& USR & USR-SR       
& 0.12            \\ 
\hline
6      
& USR-SR &USR-SR 
& 3.29 
\\ 
\hline
\end{tabular}%
}
\caption{Contributions to the numerically evaluated 1LC \eqref{1LC2} for the `Matched Peak' potential model (see Tab.~\ref{tab:parameters}) from different regions of the $(N_1,N_2)$-plane in the right panel of Fig.~\ref{fig:IntCol}. Each region is defined by the dynamics of the model at the times $N_1$ and $N_2$.}
\label{tab:intregion}
\end{table}
The numerical contributions from each region are recorded in Tab.~\ref{tab:intregion}. Note that when we evaluate Eq.~\eqref{1LC2} numerically we integrate over the whole lower triangle ($10\leq N_1 \leq 18$, $N_2\leq N_1$), but in the following discussion we don't consider any region other than the numbered ones in the right panel of Fig.~\ref{fig:IntCol}. This is because other areas of the $(N_1,N_2)$-plane correspond to when the system is in SR and less than $0.1\%$ of the total 1LC would be obtained from integrating over these areas.

Returning to the three features identified previously, we can now quantify the contribution from the first feature, A, and what it corresponds to. In the left panel of Fig.~\ref{fig:IntCol}, it can be seen that the positive contribution from this feature is spread out over the regions we've labelled 4,5 and 6 in the right panel. It is also accompanied by a negative contribution spread over just regions 4 and 5. Region 6 corresponds to the contribution of the USR-SR transition to the total 1LC and it can be seen in Tab.~\ref{tab:intregion} that it does indeed contribute most of the 1LC as expected, but not all of it. Only taking into account this transition leads to missing $23\%$ of the total 1LC.
As for the contribution of feature A in regions 4 and 5, it is around an order of magnitude smaller than the contribution from region 6.
The reduced size of the contributions from these two areas is expected when one considers what they represent in terms of the dynamics of the model. 
Region 4 corresponds to times during which the system is in USR. In a model with instantaneous transitions, you would expect no contribution from this region since $\eta'=0$ at these times, but in a potential model this is no longer the case since the $\eta'$-interaction is never `switched off'. Region 5 can be thought of as a cross-term between times during USR and during the USR-SR transition. Neglecting the contribution from these two regions would lead to an $11\%$ error in the calculated value of the 1LC. 

Feature B can be thought of as a cross-term between the SR-USR transition and other phases of evolution after the peak mode functions have been enhanced. This offers an explanation for its size. This contribution involves times during which transitions are occurring (and so $\eta'$ is large), and times at which some mode functions have already been enhanced. In Fig.~\ref{fig:Integrand} it looks as though this feature is the largest one, and might dominate the contribution to the 1LC. What we notice, however, is that this feature contains both a large positive and negative spike. This is also clear from the left panel of Fig.~\ref{fig:IntCol}. If we integrate over just this feature (so considering regions 2 and 3 in the right panel together), we find that the positive and negative spikes cancel out almost entirely. Despite it being the feature with the largest magnitude, neglecting its contribution to the 1LC would lead to a mere $1\%$ error.

The final listed feature, C, arises purely due to the initial SR-USR transition. 
In Fig.~\ref{fig:Integrand}, this feature is clearly subdominant for $k<k_{\rm peak}$, but grows in magnitude as we consider scales past $k_{\rm peak}$.
In other words, the modes for which feature C is sizeable are deep inside of the horizon at the time of the SR-USR horizon. 
Due to its growth on small scales, one might worry that if one was to extend the momentum domain of integration to $k\gg k_e> k_\text{peak}$ feature C would come to dominate the 1LC and drive its size to large values.  
We argue one should not integrate far into the plateau as in realistic models the power spectrum would exhibit a well defined peak, and modes $k\gg k_\text{peak}$ will be much less enhanced, see also the discussion in Sec.~\ref{sec: change the upper scale limit}.  
In the present analysis, feature C is contained entirely in region 1, and integrating over it yields a contribution to the 1LC a single order of magnitude smaller than the dominant feature, A. 
Neglecting the SR-USR transition would therefore yield a $10\%$ error in the value of the 1LC. 
Interestingly, this contrasts starkly with the contribution of the SR-USR transition in the instantaneous models discussed in Sec.~\ref{sec:loop with instantaneous transitions}. 
We have checked that in these models the SR-USR transition contributes $0.003\%$ of the total 1LC, showing that the assumption made in Ref.~\cite{Kristiano:2022maq} to neglect it is justified.

To summarise, the integrand, $\mathcal{I}(k,N_1,N_2)$, has a rather complicated structure, but its dominant features conform to our expectations of how the 1LC should behave. The largest contribution to the 1LC is an order of magnitude larger than any other and comes from integrating over times corresponding to the USR-SR transition. After this, the two largest contributions to the 1LC come from the initial SR-USR transition and the USR phase.
\begin{figure}
\centering
\begin{subfigure}{.49\textwidth}
  \centering
  \includegraphics[width=\linewidth]{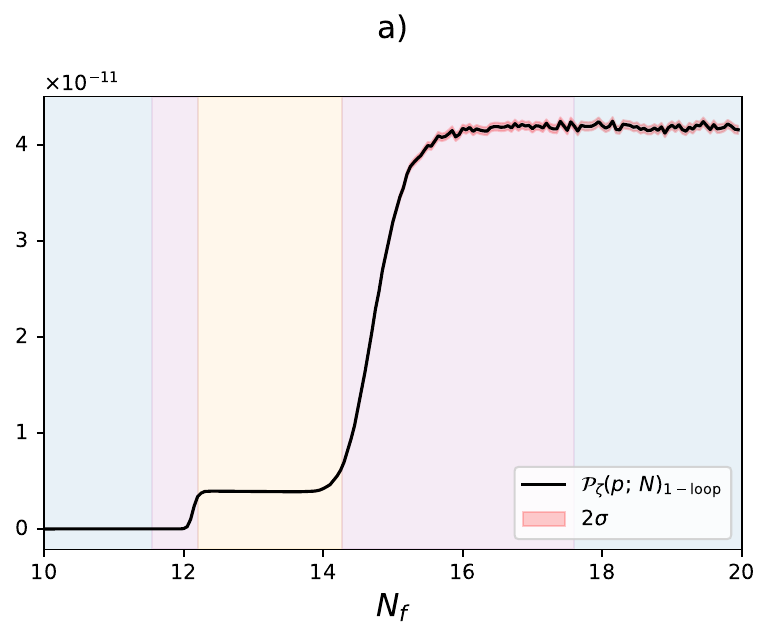}
\end{subfigure}
\begin{subfigure}{.49\textwidth}
  \centering
  \includegraphics[width=\linewidth]{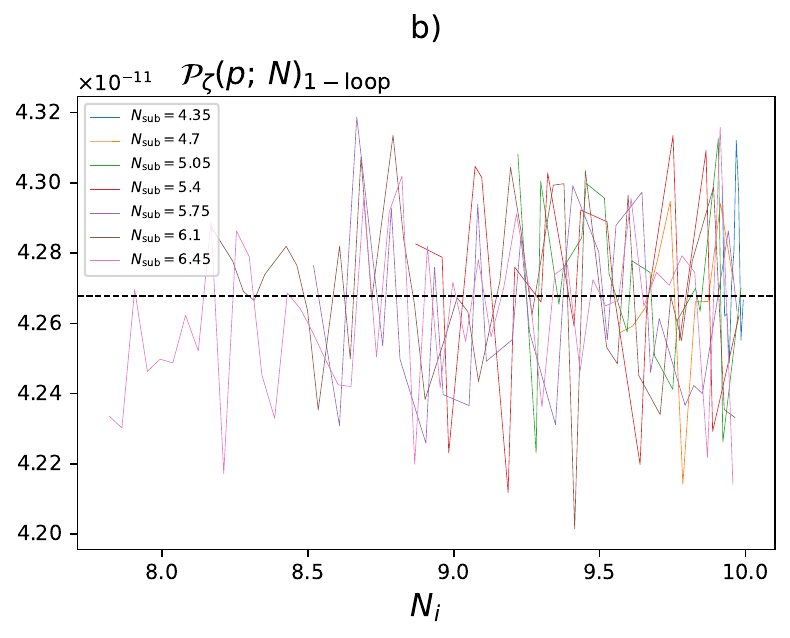}
\end{subfigure}
\begin{subfigure}{.49\textwidth}
  \centering
  \includegraphics[width=\linewidth]{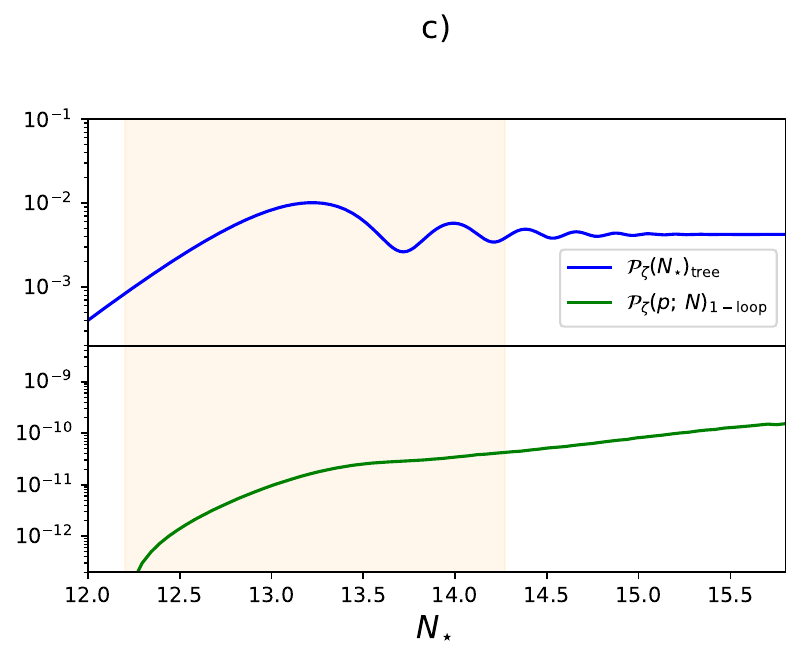}
\end{subfigure}
\begin{subfigure}{.49\textwidth}
  \centering
  \includegraphics[width=\linewidth]{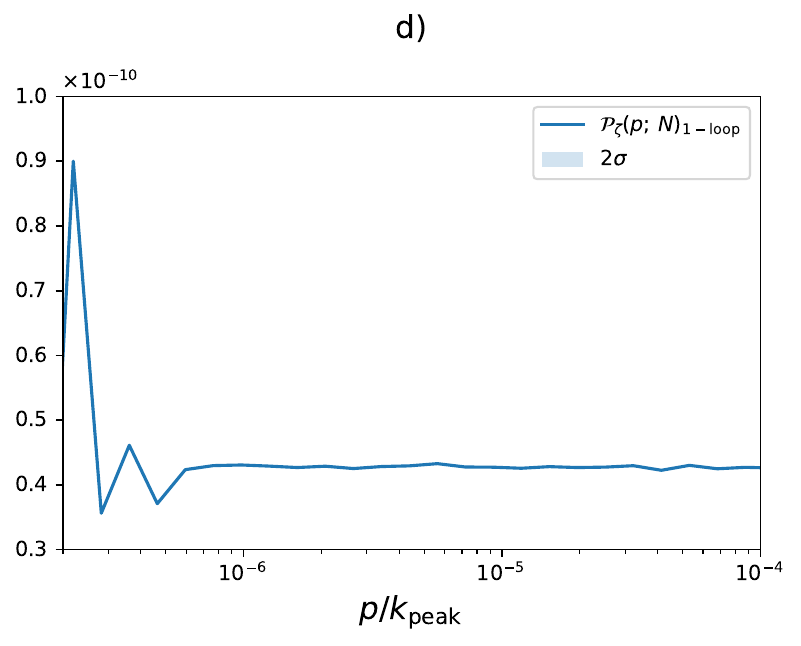}
\end{subfigure}
\caption{The quantities in all figures have been calculated using Eq.~\eqref{1LC2} for the `Matched Peak' potential model on the left of Fig.~\ref{fig:PS}. a) Value of the 1LC as a function of the upper time limit of integration, $N_f$. We set $N=35$, $N_i=10$ and $k_s$ and $k_e$ to the scales exiting the horizon at the times marked by the boundaries of the USR (orange) region. The initial conditions \eqref{MFinitial 1}--\eqref{MFinitial 2} are imposed for each mode 5 e-folds before crossing the horizon. The shaded red band displays the $2\sigma$-error estimates for the value of the 1LC according to the Monte-Carlo integrator. b) Value of the 1LC as a function of the lower time limit of integration, $N_i$. Each coloured line represents a different value of e-folds of sub-horizon evolution, $N_{\rm sub}$. c) \textit{Top}: Tree-level dimensionless power spectrum as a function of the modes horizon-crossing time, $N_\star$. \textit{Bottom}: Value of the 1LC as a function of the horizon-crossing time, $N_\star$, of the upper momentum limit of integration. The orange band indicates scales exiting during the USR phase of the model; these are the scales Eq.~\eqref{1LC2} is usually integrated over. d) Value of the 1LC as a function of the large scale, $p$. The shaded blue band displays the $2\sigma$-error estimates for the value of the 1LC according to the Monte-Carlo integrator. In our calculations, $p$ is usually separated from the peak scale by 4 orders of magnitude.}
\label{fig:Robustness}
\end{figure}

\subsection{Varying the limits of the time integrals}
\label{sec: change limits of time integration}

Treating Eq.~\eqref{1LC2} as the 1LC, we would expect its value to be robust against varying the limits of the time integration. Significant contributions to the 1LC should come only from times during which $\eta'$ is large, i.e. during transition times. Therefore, extending the time integration limits beyond the transition periods (either earlier than the SR-USR transition or later than the USR-SR transition) should not significantly affect the value of the 1LC.

\subsubsection*{Varying the upper limit of the time integral}
In panel a) of Fig.~\ref{fig:Robustness}, we investigate the effect of varying the upper limit of the time integration, $N_f$ in Eq.~\eqref{1LC2}, on the 1LC. The range of scales integrated over, the number of e-folds of sub-horizon evolution, and the lower time limit $N_i$ are all fixed. We set $N_i=10$ (a time at least an e-fold before the SR-USR transition occurs) and vary $N_f$ between 10 and 20 (a time several e-folds after the USR-SR transition has ended). Each mode is given 5 e-folds of sub-horizon evolution and we integrate over scales exiting during the USR phase, as usual.

We see in a) that all significant contributions to the 1LC come from transition periods. During these times, the value of the 1LC `steps up' as the transition progresses. When the transition comes to an end, the 1LC eventually plateaus to a fixed value at which it remains as the system enters a constant-$\eta$ phase. The fluctuations seen at late times in a) are due to the computational difficulty of the numerical integration. The $2\sigma$-error estimates from the Monte-Carlo integration are displayed as shaded bands. 
Up to numerical errors, the value of the 1LC is constant once this late-time plateau is reached.
This means that so long as the upper time limit $N_f$ is later than the end of the USR-SR transition, the 1LC is independent of its value\footnote{This is only true so long as the integration is not extended to very late times. At around $N=30$ (over 10 e-folds after the USR-SR transition), the integrand, $\mathcal{I}(k,N_1,N_2)$, begins to grow at a rate $\sim e^{2N}$. At these times, however, $\eta'$ becomes SR suppressed and we can no longer assume that $H_{\rm int}$ takes the form in Eq.~\eqref{int}, and consequently, that the 1LC is given by Eq.~\eqref{1LC2}. Physically, one would also expect inflation to end before $\mathcal{I}(k,N_1,N_2)$ begins growing. In the regime where our modelling assumptions remain valid, then, it is true that our value for the 1LC is independent of $N_f$.}.

One other point of interest is the size of the contribution to the 1LC arising from the SR-USR transition. The contribution from this transition is usually neglected (at least when working with instantaneous models) since, even though $\eta'$ is large during these times, the mode functions have not yet been enhanced. What we see from a), however, is that although the contribution from the first transition is indeed sub-dominant, it is only roughly a single order of magnitude smaller than the main contribution from the USR to SR transition. This is in line with our analysis of the contributions from various regions of the integrand in the previous section. 

\subsubsection*{Varying the lower limit of the time integral}
We can also verify that no significant contribution to the 1LC comes from integrating over early times, i.e. times before any transition has occurred. We test this by extending the lower time integration limit, $N_i$, to smaller values and represent the results in panel b) of Fig.~\ref{fig:Robustness}. Integrating from earlier times isn't as straightforward as simply altering the value of $N_i$ when we integrate. Recall that the initial conditions from Eqs.~\eqref{MFinitial 1}--\eqref{MFinitial 2} are implemented for each mode a specific number of e-folds before horizon crossing, $N_{\rm sub}$. In most of our runs, we set $N_{\rm sub}=5$. Since the final mode integrated over, $k_e$, exits the horizon at $N\approx14$, the earliest time we can consider integrating from usually is $N_i \approx 9$ since, before this time, the time evolution of the $k_e$ mode function is not defined. To integrate over earlier times, then, we must increase the value of $N_{\rm sub}$. Note that for larger $N_{\rm sub}$, this becomes computationally expensive, as the time evolution of the mode functions is highly oscillatory when the modes are deep inside the horizon.

In b), we plot the value of the 1LC as we reduce the value of $N_i$, the initial time from which we integrate. Each line represents a case with a different value of $N_{\rm sub}$. What we see from this figure is that neither increasing the length of sub-horizon evolution, $N_{\rm sub}$, nor extending the integration to earlier times has any noticeable effect on the value of the 1LC.

\subsection{Varying the upper limit of the momentum integration}
\label{sec: change the upper scale limit}

A fully rigorous computation of the 1LC would involve renormalisation and regularisation of either Eq.~\eqref{1LC1} for an instantaneous model, or Eq.~\eqref{analytic1LC1} for a potential model, as both expressions involve a divergent momentum integral. Nevertheless, we follow Ref.~\cite{Kristiano:2022maq} in obtaining a finite value for the 1LC by integrating over peak scales only. Note that, although this choice is physically motivated, the precise values of the integral limits are arbitrary.
In this section we test the robustness of the 1LC value against changes in the upper momentum limit.

From panel c) of Fig.~\ref{fig:Robustness} we see that the size of the 1LC grows as we include scales closer and closer to the peak in our integration. This growth begins to slow, however, once we've reached $k_{\rm peak}$ and does not increase significantly as we integrate up to $k_e$ -- this scale is marked by the right-hand boundary of the orange region in panel c) of Fig.~\ref{fig:Robustness}. 
The value of the 1LC at this point is the value we quote for the 1LC earlier in this work. 
If we continue to extend our integration over momenta by including scales beyond $k_e$, i.e. modes corresponding to the plateau of $\mathcal{P}_\zeta(k;N)_\text{tree}$, the 1LC continues to grow steadily.
We note that we have checked that the 1LC behaves in the same way also in instantaneous models when one integrates past $k_e$ in Eq.~\eqref{1LC1}. 
We expect that this would differ for a realistic model with a well defined peak. 
In this case, modes with $k>k_e$ would be less enhanced and therefore would probably not significantly contribute to the 1LC.

\subsection{Varying the large scale at which the 1LC is evaluated}
\label{sec: dependence of 1LC on p}
It is difficult to numerically compute Eq.~\eqref{1LC2} for a realistic CMB scale $p\sim k_{\rm CMB}=0.05\,\text{Mpc}^{-1}$. 
Indeed, a larger separation of scales would require evolving the mode functions for longer times, and this would introduce large numerical errors. 
Instead of calculating the 1LC for a CMB scale, we evaluate Eq.~\eqref{1LC2} for a scale, $p$, sufficiently removed from peak scales, but much smaller than the CMB scale, $k\gg p\gg k_{\rm CMB}$. 
One might wonder whether pushing $p$ to larger and larger scales would change the amplitude of the 1LC.  
In this section we enhance the separation between scales (as much as possible before incurring large numerical errors), and test the robustness of the 1LC. 

While in Sec.~\ref{sec:loop with instantaneous transitions} we have shown that the 1LC for an analytical model with instantaneous transitions and scale-invariant power spectrum at large scales 
is independent of $p$, this is not evident from Eq.~\eqref{1LC2}. 
In panel d) of Fig.~\ref{fig:Robustness} we show that shifting $p$ to larger and larger scales has no significant effect on the value of the 1LC as calculated from Eq.~\eqref{1LC2}. 
Note that we are able to shift $p$ approximately two orders of magnitude further away from the peak scale with no noticeable change in the value of the 1LC. 
Beyond this, however, the accuracy of our numerics breaks down and the result can no longer be trusted, as demonstrated by the widening of the $2\sigma$ numerical error. 

\section{Discussion}
\label{sec: discussion}
It was originally claimed by Kristiano \& Yokoyama in Ref.~\cite{Kristiano:2022maq} that, in single-field inflationary models capable of producing PBHs, the 1LC to the primordial power spectrum at large scales is sizeable, and endangers the perturbativity of the underlying model. 
Their conclusion was that PBH formation in single-field inflation is ruled out. 
The calculation in Ref.~\cite{Kristiano:2022maq} is set-up for an analytic model with instantaneous transitions between the SR and USR phases.
In this work, we follow the same method for calculating the 1LC as in Ref.~\cite{Kristiano:2022maq} but relax the assumption of instantaneous transitions, working instead with an inflationary model derived directly from an analytic potential. 
Previous investigations of non-Gaussianity in models with a transient USR phase might lead one to expect that working with consistent, smooth background dynamics could drastically reduce the size of the large 1LC found in Ref.~\cite{Kristiano:2022maq}. 
Indeed, while the primordial non-Gaussianity, $f_{\rm NL}$, produced from a transient USR phase can be large for models featuring instantaneous transitions between the USR and final SR eras, realistic smooth background dynamics leads to a strong suppression of the final value of $f_{\rm NL}$; see Ref.~\cite{Cai:2018dkf}. 
While this possibility was already speculated upon by other authors~\cite{Riotto:2023gpm, Firouzjahi:2023aum, Firouzjahi:2023ahg}, Franciolini \textit{et al.}~\cite{Franciolini:2023lgy} were the first to explicitly investigate it by assuming a functional form of the second Hubble SR-parameter, $\eta(N)$, which allows one
to interpolate between instantaneous and smooth transitions. 
Our set-up differs from the one of Ref.~\cite{Franciolini:2023lgy} in that we derive the smooth background dynamics and the evolution of the curvature perturbation  directly from a specific potential. Like Ref.~\cite{Franciolini:2023lgy}, we then proceed to compute the resulting 1LC numerically.

One difference between our approach and
that of Ref.~\cite{Franciolini:2023lgy} is that we compare potential and instantaneous models producing dimensionless power spectra with the same peak amplitude, not directly possible when one assumes a functional form for $\eta(N)$ \cite{Cole:2022xqc, Franciolini:2023lgy}. 
Indeed, 
smoothing the transition in $\eta(N)$ will change the integrated duration of the USR phase and hence the amplitude of the peak. 

By comparing two models, one featuring instantaneous transitions and the other produced from a potential, which yield the same peak amplitude relevant for PBH production, $\mathcal{P}_\zeta(k_{\rm peak})_{\rm tree}=0.01$, we find that the 1LCs induced on large sales differ by 26\% only -- less than one order of magnitude.
In other words, by fixing the feature that PBH production is most sensitive to, $\mathcal{P}_\zeta(k_{\rm peak})_{\rm tree}$, the size of the 1LC is largely unaffected by whether the model includes instantaneous transitions or is derived from a well-defined potential. 

Nevertheless, we do not agree with the original conclusion of Ref.~\cite{Kristiano:2022maq} that PBH formation in single-field inflation is ruled out. 
Upon numerically evaluating our expressions for the 1LC for both an instantaneous and potential model capable of producing PBHs, we find values for their 1LCs nearly two orders of magnitude smaller than the tree-level amplitude of their spectra at large scales. 
While this 1LC may be large enough to cause some concern, it is not at the level sufficient to question perturbativity. 
We find that the discrepancy between our conclusion and that of Ref.~\cite{Kristiano:2022maq} lies in our full consideration of the transition of the model back to SR; see Sec.~\ref{sec:loop instantaneous model}.
Thus, a single-field inflationary model capable of producing PBHs can be constructed whose 1LC is consistent with perturbativity. 

An additional advantage of our numerical approach  
is that it enables us to perform a variety of robustness checks of our results.
The values we obtain for the 1LC behave as expected under changes of the various integration limits in Eq.\eqref{1LC2}, solidifying our confidence that our results
are accurate, assuming the validity of the underlying In-In computation.

We stress again that our approach follows the same basic framework as in Ref.~\cite{Kristiano:2022maq}, and that there is considerable debate about whether this includes all relevant contributions to the 1LC, and is robust in the light of  regularization and renormalisation, see e.g. Refs.~\cite{Fumagalli:2023hpa,Firouzjahi:2023aum, Firouzjahi:2023bkt}. In particular, our analysis only includes the contribution to the 1LC due to cubic interactions, see Fig.~\ref{fig:Diagrams}. Quartic interactions contribute as well at one-loop level, see Refs.~\cite{Firouzjahi:2023aum, Firouzjahi:2023bkt} for a recent analysis. We hope to extend our numerical analysis to the study of 1LC mediated by quartic interactions in the future.

Nevertheless, it was important to understand how the size of the 1LC as calculated in Ref.~\cite{Kristiano:2022maq} would be affected by working from an analytic potential. 
Since we find very minimal difference, we can conclude that relaxing the assumption of instantaneous transitions does not significantly reduce the size of the 1LC. While we expect our result to be unaffected by the specific form of the potential chosen to model a smooth SR to USR to SR dynamics, we note that the toy model we employ yields a plateau of enhanced modes. The motivation for this choice is that the potential \eqref{potential} mimics closely the dynamics and perturbations produced in the instantaneous models of Ref.~\cite{Kristiano:2022maq}, see Fig.~\ref{fig:PS}. Clearly, our work should now be extended to the study of the 1LC produced from generic single-field potentials, yielding a more realistic power spectrum with a reduction in amplitude at scales smaller than the peak. As suggested in Refs.~\cite{Riotto:2023hoz,Tada:2023rgp,Firouzjahi:2023bkt}, the modes contributing to the rise and fall of the peak could potentially reduce the 1LC. We leave this for future work.

\section*{Acknowledgments}
It is a pleasure to thank David Seery for many interesting conversations on the topic of loop corrections. 
MWD is supported by a studentship awarded by the Perren Bequest.  LI was supported by a Royal Society
funded postdoctoral position for much of this work and acknowledges current
financial support from the STFC under grant ST/X000931/1.
DJM is supported by a Royal Society University Research Fellowship.

\bibliography{refs} 
\bibliographystyle{JHEP}

\end{document}